\documentclass[journal,twoside]{IEEEtran}

\IEEEpubid{\begin{minipage}{\textwidth}\ \\[12pt] \centering
  1549-8328 \copyright 2016 IEEE. Personal use is permitted, but republication/redistribution requires IEEE permission.\\
  See http://www.ieee.org/publications standards/publications/rights/index.html for more information.
\end{minipage}}

\usepackage{amssymb,amsmath,amsthm,bm}
\usepackage{diagbox}
\usepackage{subfigure}
\newtheorem{lemma}{Lemma}
\newtheorem{theorem}{Theorem}
\usepackage{multirow,bigstrut}
\usepackage{tabularx}
\usepackage{url}
\usepackage{array}
 \usepackage{color}
\newcolumntype{x}[1]{>{\centering\arraybackslash\hspace{0pt}}p{#1}}
\newcolumntype{C}[1]{>{\centering}p{#1}}

\usepackage[normalem]{ulem}

\usepackage{graphicx}
\graphicspath{{figures-pdf/}}

\DeclareMathOperator{\diag}{diag}   % diagonal matrix

\newlength\figwidth
\newlength\sfigwidth
\begin{document}

\title{Theoretical Design and FPGA-Based Implementation of Higher-Dimensional Digital Chaotic Systems}
	
\author{Qianxue Wang, Simin Yu, Chengqing Li, Jinhu L\"u, Xiaole Fang, Christophe Guyeux, and Jacques M. Bahi
% <-this % stops a space
\thanks{Manuscript received October 9, 2015; revised December 7, 2015; accepted
December 21, 2015. Date of publication March 30, 2016; date of current
version April 5, 2016. This work was supported by the National Science and
Technology Major Project of China under Grant 2014ZX10004001-014; by
the National Natural Science Foundation of China under Grants 61532020,
61172023, and 11472290; by the China Postdoctoral Science Foundation under
Grant 2014M552175; by the Scientific Research Foundation for the Returned
Overseas Chinese Scholars, State Education Ministry; by the Hunan Provincial
Natural Science Foundation of China Grant 2015JJ1013. This paper was
recommended by Associate Editor J. M. Olm.}
\thanks{Q. Wang and S. Yu are with College of Automation, Guangdong University of Technology, Guangzhou 510006, Guangdong, China.}
\thanks{C. Li is with College of Information Engineering, Xiangtan University, Xiangtan 411105, Hunan, China (e-mail: DrChengqingLi@gmail.com).}
\thanks{J. L\"u is with Academy of Mathematics and Systems Sciences, Chinese Academy of Sciences, Beijing 100190, China.}
\thanks{X. Fang is with Land and Resources Technology Center of Guangdong Province, Guangzhou 510075, Guangdong, China.}
\thanks{C. Guyeux and J. M. Bahi are with Femto-st Institute, UMR 6174 CNRS, University of Franche-Comt\'{e}, Besan\c{c}on 25000, France.}
\thanks{Copyright (c) 2015 IEEE. Personal use of this material is permitted. However, permission to use this material for any other purposes must be obtained from the IEEE by sending an email to pubs-permissions@ieee.org.}
}
	
\markboth{IEEE TRANSACTIONS ON CIRCUITS AND SYSTEMS--I: REGULAR PAPERS, VOL. 63, NO. 3, MARCH 2016}{Yu \MakeLowercase{\textit{et al.}}:}
		
\maketitle

\begin{abstract}
Traditionally, chaotic systems are built on the domain of infinite precision in mathematics. However, the quantization is
inevitable for any digital devices, which causes dynamical degradation. To cope with this problem, many methods were proposed,
such as perturbing chaotic states and cascading multiple chaotic systems. This paper aims at developing a novel methodology to
design the higher-dimensional digital chaotic systems (HDDCS) in the domain of finite precision. The proposed system is based
on the chaos generation strategy controlled by random sequences. It is proven to satisfy the Devaney¡¯s definition of chaos. Also, we
calculate the Lyapunov exponents for HDDCS. The application of HDDCS in image encryption is demonstrated via FPGA platform.
As each operation of HDDCS is executed in the same fixed precision, no quantization loss occurs. Therefore, it provides a perfect
solution to the dynamical degradation of digital chaos.
\end{abstract}
	
% \fi
% For peer review papers, this IEEEtran command inserts a page break and
% creates the second title.
\IEEEpeerreviewmaketitle

\section{Introduction}

\IEEEPARstart{C}{haotic} system of simple structure can demonstrate complex dynamical properties in infinite mathematical world,
such as sensitivity to initial conditions, topological transitivity and mixing, expansiveness, and decaying autocorrelation function
\cite{MAY:Logistic:Nature1976}. Such properties have subtle relation with some requirements of secure encryption system, especially
sensitivity with respect to change of secret key~\cite{YaobinMao:CSF2004,LiShujun:Rules:IJBC2006,kanso2009logistic}. So, designing chaos-based encryption schemes emerged as a new research direction to reinforce information security of data sent through the Internet~\cite{Kohda:Statistics:TIT1997,Gregory:Commun:Science98,Lee:lyapunov:CPC2004,Kwok:finiteprecision:2007,Cho:QKD:TCASI15}. However, any digital chaotic system implemented in finite-precision devices definitely degrade of various extents \cite{beck1987effects,blank1994pathologies}, which may cause serious security flaws \cite{Binder:Logistic:PRA86,Lisj:precision:CPC2003,HPhu:DCS:SMCS2015}. Multiplication, division, and other linear/nonlinear operations of real numbers may incur error by rounding, quantization and/or overflow,
which may further result in large differences between theoretical chaotic orbits and actual ones~\cite{Galias:Round:CASM13,Deng:hybrid:IS2015}. To counteract such degradation, many methods were proposed, such as perturbing chaotic states\cite{hong1997realizing,tao1998perturbance}, perturbing control parameter \cite{Cernak:DCS:PLA1996}, \cite{Sang:Perturbance:EL1998}, \cite{LiCY:PRNS:VLSI2012}, or cascading two and more chaotic maps \cite{wolff1986transients,Nagaraj:EPJT:2008,YCZhou:Switching:TCASI2014}. Recently, Deng and Hu showed that phase space of chaotic system must not be implemented in infinite precision, i.e., the definitions of chaos, like Devaney's chaos, are compatible with domain of finite precision~\cite{HPhu:DCS:SMCS2015}.

\IEEEpubidadjcol
%  must call \IEEEpubidadjcol in the second column for its text to clear the %IEEEpubid mark.

Some works were done to disclose degradation characteristic of some simple digital domain chaotic maps, such as Logistic map \cite{Li:logistic:ND2014} and
piecewise linear chaotic map \cite{Li:DPWLCM:IJBC2005}. But, there is still lack of systematic theory to rigorously analyze and overcome
degradation caused by finite precision effects. In 2010, Bahi \textit{et al.} proposed an 1-D integer domain chaotic system (IDCS) designated as Chaotic Iterations (CI) system \cite{Guyeux:Hash:JACT10}. The CI system receives one or two random stream(s) as input and mixes them with chaotic iterations, where
the involved mixture function is only bitwise OR operation. It was theoretically proven that the proposed systems satisfy the Devaney's definition of chaos
given in \cite{Devaney:Chaos:2003}. Since the systems run on finite sets of integer domains, the above problem caused by finite-precision disappears.
As there is no any transformation from real numbers to binary sequences, CI system can be considered as one of the most effective solutions for
solving the degradation problem of digital chaotic systems.

In \cite{Bahi:XORshift:JNCA04}, a mark sequence was applied to avoid wasteful duplication of values, leading to an obvious speed improvement.
In \cite{Bahi:PRNS:IJAS11}, chaotic combination of two input entropic streams has solved flaws exhibited in the system designed in \cite{Bahi:XORshift:JNCA04}.
The chaos generation strategy was implemented in~\cite{SMYu:Chaotifying:IJBC2012,SMYu:integer:IJBC2014,SYu:ARM:CASVT15}, through a sample-hold
circuit and a decoder circuit so as to convert the uniform noise signal into a random sequence. The second chaos generation strategy named chaotic
bitwise dynamical system (CBDS) was proposed in~\cite{Wang:bitwise:CPB15}. All the previous related works deal with 1-D chaotic maps,
the counterexamples in higher-dimensional domain and their dynamical properties are still unknown to date.

In this paper, we study the general problem of constructing higher-dimensional digital chaotic system (HDDCS) in digital devices with finite precision, and establish general framework of composing HDDCS. We demonstrate that the associated state network of HDDCS is strongly connected, and then further prove that HDDCS satisfies the Devaney's definition of chaos in the domain of finite precision. In addition, Lyapunov exponents for such a HDDCS are also given, and FPGA-based implementations are then provided to show the potential application of HDDCS in the digital world. The basic advantage and practical benefits of the proposed approach is that the chaos satisfying Devaney's definition can be generated in the real digital devices by means of the chaos generation strategy controlled by random sequences.

The remainder of this paper is organized as follows. The description of HDDCS is given in Sec.~\ref{DCS_Strong}, while proof of
its chaos properties is provided in Sec.~\ref{ChaosProofHDDCS}. The Lyapunov exponents of HDDCS are calculated in Sec.~\ref{Lyapunovexponent}.
Section~\ref{application} presents application of HDDCS in image encryption with FPGA implementation. The last section concludes the paper.

\section{Design of HDDCS}
\label{DCS_Strong}

\subsection{Description of HDDCS}

The general form of iterative equation for $m$-dimensional digital system with $N$-bit fixed precision is
\begin{equation*}
\left\{
\begin{IEEEeqnarraybox}[][c]{ll}
\IEEEstrut
x_1^n & = F_1(x^{n-1}_{1},x^{n-1}_{2},\ldots, x^{n-1}_{m}),\\
x_2^n & = F_2(x^{n-1}_{1},x^{n-1}_{2},\ldots, x^{n-1}_{m}),\\
	  &\vdots \\
x_m^n & = F_m(x^{n-1}_{1},x^{n-1}_{2},\ldots, x^{n-1}_{m}),
\IEEEstrut
\end{IEEEeqnarraybox} \right.
\end{equation*}
where $F_1$, $F_2$, $\cdots$, $F_m$ are some iterative functions, $x_1^n,x_2^n, \cdots, x_m^n$ and $x_1^{n-1},x_2^{n-1}, \cdots, x_m^{n-1}$ can be represented as the following binary form:
\begin{equation}
\left\{
\begin{IEEEeqnarraybox}[][c]{ll}
\IEEEstrut
x_1^n    & = x^{n}_{1,P-1}x^{n}_{1,P-2}\ldots x^{n}_{1,0}.x^{n}_{1,-1}x^{n}_{1,-2}\ldots x^{n}_{1,-Q},\\
x_1^{n-1}& = x^{n-1}_{1,P-1}x^{n-1}_{1,P-2}\ldots x^{n-1}_{1,0}.x^{n-1}_{1,-1}x^{n-1}_{1,-2}\ldots x^{n-1}_{1,-Q},\\
x_2^n    & = x^{n}_{2,P-1}x^{n}_{2,P-2}\ldots x^{n}_{2,0}.x^{n}_{2,-1}x^{n}_{2,-2}\ldots x^{n}_{2,-Q},\\
x_2^{n-1}& = x^{n-1}_{2,P-1}x^{n-1}_{2,P-2}\ldots x^{n-1}_{2,0}.x^{n-1}_{2,-1}x^{n-1}_{2,-2}\ldots x^{n-1}_{2,-Q},\\
         & \vdots \\
x_m^n    & = x^{n}_{m,P-1}x^{n}_{m,P-2}\ldots x^{n}_{m,0}.x^{n}_{m,-1}x^{n}_{m,-2}\ldots x^{n}_{m,-Q},\\
x_m^{n-1}& = x^{n-1}_{m,P-1}x^{n-1}_{m,P-2}\ldots x^{n-1}_{m,0}.x^{n-1}_{m,-1}x^{n-1}_{m,-2}\ldots x^{n-1}_{m,-Q},
\IEEEstrut	
\end{IEEEeqnarraybox}\right.
\label{eq:binary}
\end{equation}
and $P+Q=N$.
	
Set the general expression of $m$ one-sided infinite random sequences as
\begin{equation}
\left\{
\begin{IEEEeqnarraybox}[][c]{ll}
\IEEEstrut
	s & = s^1s^2\ldots s^n\ldots\\
	u & = u^1u^2\ldots u^n\ldots\\
	  & \vdots\\
	v & = v^1v^2\ldots v^n\ldots
\IEEEstrut
\end{IEEEeqnarraybox}
\right.
\label{eq:side}
\end{equation}

Likewise, each random number of these $m$ sequences is expressed in binary form as
\begin{equation*}
\left\{
\begin{IEEEeqnarraybox}[][c]{ll}
\IEEEstrut
	s^n & = s^{n}_{P-1}s^{n}_{P-2}\ldots s^{n}_{0}.s^{n}_{-1}s^{n}_{-2}\ldots s^{n}_{-Q},\\
	u^n & = u^{n}_{P-1}u^{n}_{P-2}\ldots u^{n}_{0}.u^{n}_{-1}u^{n}_{-2}\ldots u^{n}_{-Q},\\
	    & \vdots\\
	v^n & = v^{n}_{P-1}v^{n}_{P-2}\ldots v^{n}_{0}.v^{n}_{-1}v^{n}_{-2}\ldots v^{n}_{-Q},
\IEEEstrut
\end{IEEEeqnarraybox}
\right.
\end{equation*}
where $n\in \mathbb{Z}^+$, $\forall\ j\in \{P-1,P-2,\ldots,0,-1,-2,\ldots,-Q\}$, $s^n_j$, $u^n_j$, and $v^n_j$ are the $j$-th bit of the binary form of $s^n$, $u^n$, and $v^n$, respectively. Note that the range of random numbers satisfy $s^n, u^n, \ldots, v^n\in [0,2^P-2^{-Q}]$ due to limited accuracy of $N$-bit representation.

Through the iterative update mechanism controlled by random sequences, the general form of $m$-dimensional digital chaotic system ($m$-D DCS) is
\begin{equation}
\left\{
\begin{IEEEeqnarraybox}[\IEEEeqnarraystrutmode\IEEEeqnarraystrutsizeadd{2pt}{2pt}][c]{rCl}
\IEEEeqnarraymulticol{3}{l}{x^{n}_{1,P-1}x^{n}_{1,P-2}\ldots x^{n}_{1,0}.x^{n}_{1,-1}x^{n}_{1,-2}\ldots x^{n}_{1,-Q}}\\
& = & F_1(\cdot)_{s^n_{P-1}} \ldots F_1(\cdot)_{s^n_0}.F_1(\cdot)_{s^n_{-1}} F_1(\cdot)_{s^n_{-2}}\ldots F_1(\cdot)_{s^n_{-Q}},\\
\IEEEeqnarraymulticol{3}{l}{ x^{n}_{2,P-1}x^{n}_{2,P-2}\ldots x^{n}_{2,0}.x^{n}_{2,-1}x^{n}_{2,-2}\ldots x^{n}_{2,-Q} }\\
& = & F_2(\cdot)_{u^n_{P-1}}  \ldots F_2(\cdot)_{u^n_0}.F_2(\cdot)_{u^n_{-1}}F_2(\cdot)_{u^n_{-2}}\ldots F_2(\cdot)_{u^n_{-Q}},\\
&\vdots & \\
\IEEEeqnarraymulticol{3}{l}{ x^{n}_{m,P-1}x^{n}_{m,P-2}\ldots x^{n}_{m,0}.x^{n}_{m,-1}x^{n}_{m,-2}\ldots x^{n}_{m,-Q} }\\
& = & F_m(\cdot)_{v^n_{P-1}}  \ldots F_m(\cdot)_{v^n_0}.F_m(\cdot)_{v^n_{-1}}F_m(\cdot)_{v^n_{-2}}\ldots F_m(\cdot)_{v^n_{-Q}},
\end{IEEEeqnarraybox}
\right.
\label{BVN12}
\end{equation}
where
\begin{equation*}
\left\{
\begin{IEEEeqnarraybox}[][c]{lll}
\IEEEstrut
	F_1(\cdot)_i & =     & F_1(x^{n-1}_{1},x^{n-1}_{2},\ldots, x^{n-1}_{m})_i\\
	F_2(\cdot)_i & =     & F_2(x^{n-1}_{1},x^{n-1}_{2},\ldots, x^{n-1}_{m})_i\\
	             &\vdots & \\
	F_m(\cdot)_i & =     & F_m(x^{n-1}_{1},x^{n-1}_{2},\ldots, x^{n-1}_{m})_i
\IEEEstrut
\end{IEEEeqnarraybox}%
\right.
\end{equation*}
denote the $i$-th component of the iterative function, where $i=P-1,P-2,\ldots,0,-1,-2,\ldots,-Q$.
	
In Eq.~(\ref{BVN12}), we define chaos generation strategy controlled by random sequences for HDDCS as
\begin{equation}
\left\{
\begin{IEEEeqnarraybox}[][c]{llll}
\IEEEstrut
x^n_{1,j}&=F_1(\cdot)_{s^n_j}&=&\left\{
\begin{IEEEeqnarraybox}[][c]{l?l}
	F_1(\cdot)_{j} &\text{ if } s^n_j=1,\\
	x^{n-1}_{1,j} &\text{ if } s^n_j=0,
\end{IEEEeqnarraybox}%
\right.\\
x^n_{2,j}&=F_2(\cdot)_{u^n_j}&=&\left\{
\begin{IEEEeqnarraybox}[][c]{l?l}
	F_2(\cdot)_{j} &\text{ if } u^n_j=1,\\
	x^{n-1}_{2,j} &\text{ if } u^n_j=0,
\end{IEEEeqnarraybox}%
\right.\\
&&\vdots&\\ x^n_{m,j}&=F_m(\cdot)_{v^n_j}&=&\left\{
\begin{IEEEeqnarraybox}[][c]{l?l}
	F_m(\cdot)_{j} &\text{if } v^n_j=1,\\
	x^{n-1}_{m,j} &\text{if } v^n_j=0,
\end{IEEEeqnarraybox}%
\right.
\IEEEstrut
\end{IEEEeqnarraybox}%
\right.
\label{eq:strategy}
\end{equation}
where $j=P-1,P-2,\ldots,0,-1,-2,\ldots,-Q$.

Then, Eq.~(\ref{BVN12}) is further expressed as the following general form:
\begin{equation}
\left\{
\begin{IEEEeqnarraybox}[][c]{lll}
\IEEEstrut
	x_1^n & =     & (x_1^{n-1}\cdot\overline{s^n})+(F_1(\cdot)\cdot s^n),\\
	x_2^n & =     & (x_2^{n-1}\cdot\overline{u^n})+(F_2(\cdot)\cdot u^n),\\
	      &\vdots & \\
	x_m^n & =     & (x_m^{n-1}\cdot\overline{v^n})+(F_m(\cdot)\cdot v^n),
\IEEEstrut
\end{IEEEeqnarraybox}%
\right.
\label{BVN2}
\end{equation}
where the operators $``\cdot"$, $``\overline{(\cdot)}"$, and $``+"$ denote bitwise AND, bitwise NOT (negation), and bitwise OR, respectively.

Let us define %metric space $(\mathcal{E}, G_F)$, where
$\mathcal{E}$ as the set of points $E$ of form $((s,u,\ldots,v),(x_1,x_2,\ldots,x_m))$,
where $s,u,\ldots, v$ are $m$ independent random sequences and $x_1$, $x_2$, $\ldots$, $x_m$ are $N$-bit real numbers. Consider a metric space $(\mathcal{E}, d)$ and a continuous function $G_F:\mathcal{E}\rightarrow\mathcal{E}$ as
\begin{equation}
\begin{IEEEeqnarraybox}[][c]{lll}
\IEEEstrut
G_F(E) & = & G_F((s,u,\ldots,v),(x_1,x_2,\ldots,x_m))\\
       & = & ((\sigma(s),\sigma(u),\ldots,\sigma(v)),(H_{F_1}(i(s),(x_1,x_2,\ldots,\\
       &   & x_m)),H_{F_2}(i(u),(x_1,x_2,\ldots,x_m)),\ldots,H_{F_m}(i(v),\\
       &   &(x_1,x_2,\ldots,x_m))),
\IEEEstrut
\end{IEEEeqnarraybox}
\label{GF}
\end{equation}
where
$\sigma(w)$ ($w\in\{s, u, \cdots, v\}$) shifts one integer in the one-sided infinite sequence $w=(w^1 w^2\ldots w^n\ldots)$ to the left,
and
\begin{equation}
\left\{
\begin{IEEEeqnarraybox}[][c]{lll}
\IEEEstrut
H_{F_1}(i(s),(x_1,x_2,\ldots,x_m))& = &((x_1\cdot\overline{i(s)})+(F_1(\cdot)\cdot i(s))),\\
H_{F_2}(i(u),(x_1,x_2,\ldots,x_m))& = &((x_2\cdot\overline{i(u)})+(F_2(\cdot)\cdot i(u))),\\
                                  & \vdots &\\
H_{F_m}(i(v),(x_1,x_2,\ldots,x_m))& = &((x_m\cdot\overline{i(v)})+(F_m(\cdot)\cdot i(v))).
\IEEEstrut
\end{IEEEeqnarraybox}
\right.
\label{BVN3}
\end{equation}

\begin{figure}[!htb]
\centering
\includegraphics[width=0.9\figwidth]{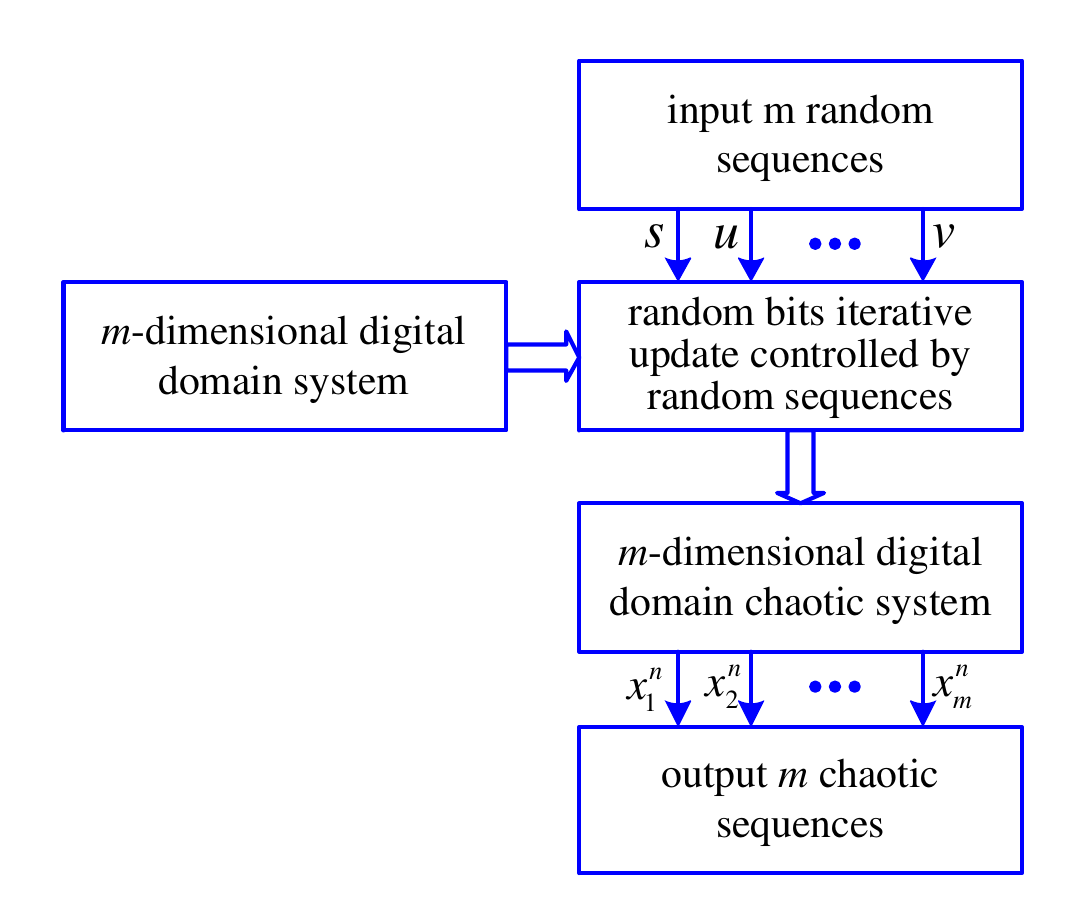}
\caption{The flowchart of the proposed higher-dimensional digital chaotic system.}
\label{Therelationship}
\end{figure}

As for function $\sigma(w)$, one has
\begin{equation*}
\sigma^k(w)=w^{k+1}w^{k+2}\ldots w^n,
\end{equation*}
where $k$ is a positive integer, and
\begin{equation*}
\sigma^k(w)\triangleq\underbrace{\sigma\circ\sigma\circ\ldots\circ\sigma}\limits_{k}(w).
\end{equation*}
Then, one can further get
\begin{equation*}
i(w)=w^k ,
\end{equation*}
where  $w\in\{s, u, \cdots, v\}$, $k\in  \mathbb{Z}^+$ and $i(w)$ is equal to the overflow from the left shifting of the sequence $w$. In other words, some bits are randomly updated by $F_1$ in each updating iteration, which are determined by $i(s)$.
Similarly, $i(u)$ determines the specific bits which are updated by $F_2$, and $i(v)$ determines the specific bits that are updated by $F_m$.
The relation among $m$-dimensional digital domain system, $m$ random sequences and $m$-dimensional digital chaotic system is shown in Fig.~\ref{Therelationship}. Note that true random sequences usually cannot be obtained via digital simulation \cite{hirano2010fast,Kocarev:TrueRandom:TCASI11}. But the input $m$ random sequences can be generated by true random number generator (TRNG) based on digital circuit artifacts, such as metastable circuits \cite{Michael:TRNG:Lecture}, and digital oscillator rings \cite{Sunar:TRNG:IEEE}.

Let $E^0=((s, u, \ldots, v),(x_1^0, x_2^0, \ldots, x_m^0))\in \mathcal{E}$ be the initial condition,
\begin{equation*}
E^k=((\sigma^{k}(s),\sigma^{k}(u),\cdots,\sigma^{k}(v))),(x_1^k,x_2^k,\cdots,x_m^k))\in \mathcal{E},
\end{equation*}
and
\begin{multline*}
E^{k+1}=((\sigma^{k+1}(s),\sigma^{k+1}(u),\cdots,\sigma^{k+1}(v)),\\
(x_1^{k+1},x_2^{k+1},\cdots,x_m^{k+1}))\in \mathcal{E}
\end{multline*}
denote the $k$-th and $(k+1)$-th iteration, respectively.

With the above notations, HDDCS can be defined as
\begin{equation*}
E^{k+1}=G_F(E^k),
\end{equation*}
where $k=0,1,2,3,\ldots$.

Let
\begin{equation*}
\left\{
\begin{IEEEeqnarraybox}[][c]{ll}
\IEEEstrut
	\hat{s} & = \hat{s}^1\hat{s}^2\ldots \hat{s}^n\ldots\\
	\hat{u} & = \hat{u}^1\hat{u}^2\ldots \hat{u}^n\ldots\\
	  & \vdots\\
	\hat{v} & = \hat{v}^1\hat{v}^2\ldots \hat{v}^n\ldots
\IEEEstrut
\end{IEEEeqnarraybox}
\right.
\end{equation*}
denote another one-sided infinite random sequences. Then, we introduce the distance $d$ in the metric space $(\mathcal{E}, d)$ as
\begin{multline*}
d(((s,u,\ldots,v),(x_1,x_2,\cdots,x_m)),\\
((\hat{s},\hat{u},\ldots,\hat{v}),(\hat{x}_1,\hat{x}_2,\ldots,\hat{x}_m)))=d_s(s,\hat{s})+d_u(u,\hat{u})+\cdots+\\
d_v(v,\hat{v})+d_{x}((x_1,x_2,\cdots, x_m),
(\hat{x}_1,\hat{x}_2, \cdots,\hat{x}_m)),
\end{multline*}
where
\begin{equation*}
\left\{
\begin{IEEEeqnarraybox}[][c]{lll}
\IEEEstrut
	d_s(s,\hat{s})&=&\sum_{k=1}^\infty \frac{|s^k-\hat{s}^k|}{2^{Nk}},\\
	d_u(u,\hat{u})&=&\sum_{k=1}^\infty \frac{|u^k-\hat{u}^k|}{2^{Nk}},\\
	              &\vdots&\\
	d_v(v,\hat{v})&=&\sum_{k=1}^\infty \frac{|v^k-\hat{v}^k|}{2^{Nk}},
\IEEEstrut
\end{IEEEeqnarraybox}%
\right.
\end{equation*}
\begin{multline*}
d_{x}((x_1, x_2, \cdots, x_m), (\hat{x}_1, \hat{x}_2, \cdots, \hat{x}_m ))\\
= \sqrt{(x_1-\hat{x}_1)^2+(x_2-\hat{x}_2)^2+\ldots+(x_m-\hat{x}_m)^2},
\end{multline*}
$x_1,\hat{x}_1,x_2,\hat{x}_2,\ldots,x_m,\hat{x}_m$ are binary forms of real numbers with $N$-bit finite precision, and $0\leq d_{x}\leq\sqrt{m}(2^P-2^{-Q}))$.
	
\subsection{Comparative study of RDCS, IDCS, CBDS, and HDDCS}

The proposed system HDDCS is briefly compared with RDCS, IDCS in~\cite{SMYu:integer:IJBC2014} and CBDS in~\cite{Wang:bitwise:CPB15},
in Table~\ref{CSDH}. More detailed differences are summarized below:
\begin{table*}[!t]
\centering
\caption{Comparison of the digital chaotic systems, RDCS, IDCS, CBDS, and HDDCS.}
\includegraphics[height=0.5\textwidth]{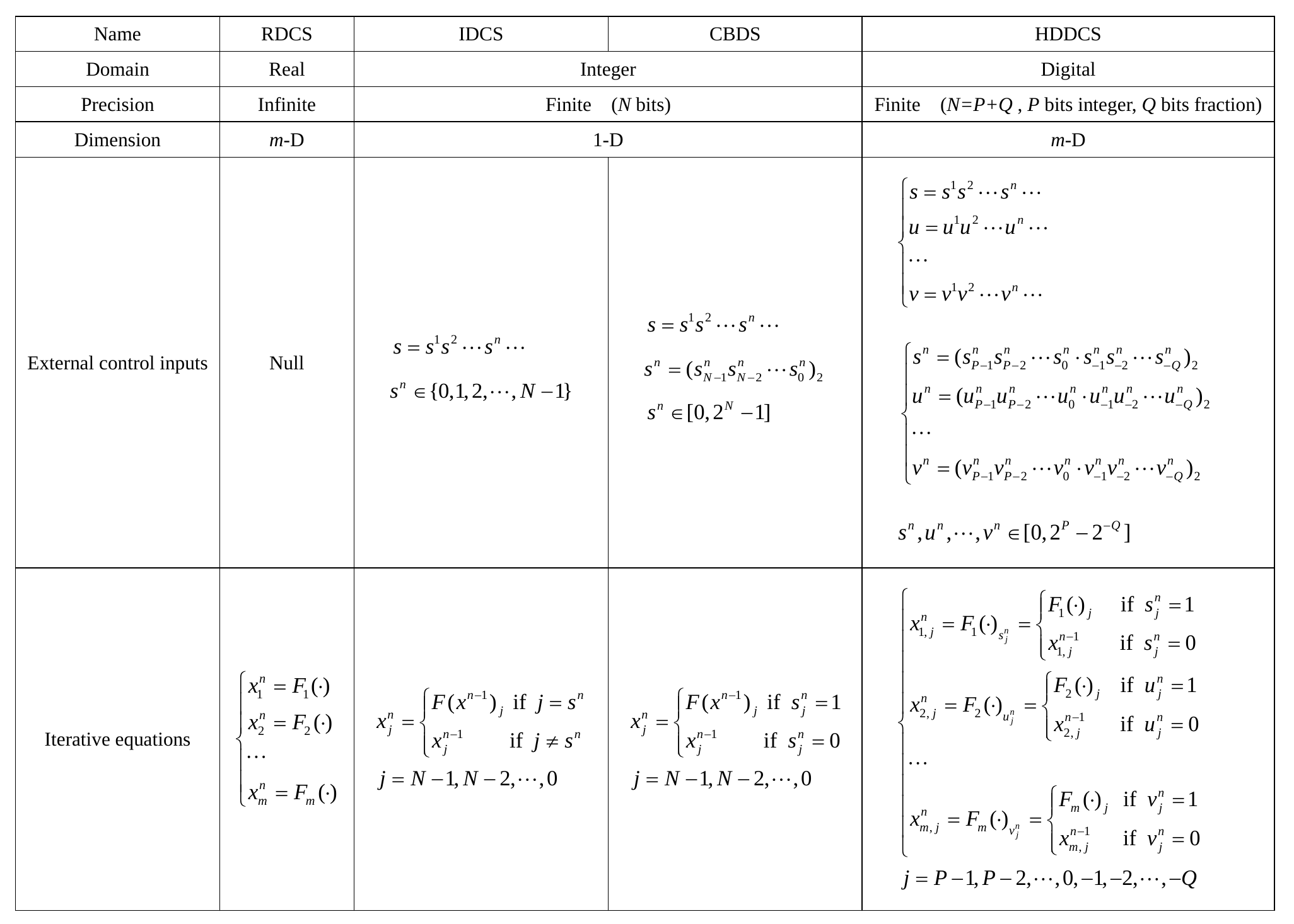}
\label{CSDH}	
\end{table*}	
\begin{itemize}
\item The adopted implementation environment of RDCS is an infinite precision. However, IDCS, CBDS, and HDDCS are implemented in finite precision,
they are suitable for digital computers or other digital devices.

\item RDCS present the finite integer part and infinite fractional part of a real number as follows:
\begin{equation*}
\left\{
\begin{IEEEeqnarraybox}[][c]{ll}
\IEEEstrut
x^n_1= & x^{n}_{1,P-1}\ldots x^{n}_{1,0}.x^{n}_{1,-1}x^{n}_{1,-2}\ldots x^{n}_{1,-Q}\ldots\\
x^n_2= & x^{n}_{2,P-1}\ldots x^{n}_{2,0}.x^{n}_{2,-1}x^{n}_{2,-2}\ldots x^{n}_{2,-Q}\ldots\\
       & \vdots \\
x^n_m= & x^{n}_{m,P-1}\ldots x^{n}_{m,0}.x^{n}_{m,-1}x^{n}_{m,-2}\ldots x^{n}_{m,-Q}\ldots	
\IEEEstrut
\end{IEEEeqnarraybox}
\right.
\end{equation*}

But IDCS and CBDS process data in integer domain as
\begin{equation*}
 x^n=x^{n}_{N-1}x^{n}_{N-2}\ldots x^{n}_{0}.
 \end{equation*}
Furthermore, in HDDCS, the range of the input data are extended to digital domain of finite integer part and finite fractional part as 	
\begin{equation*}
\left\{
\begin{IEEEeqnarraybox}[][c]{ll}
\IEEEstrut
x^n_1&=x^{n}_{1,P-1}x^{n}_{1,P-2}\ldots x^{n}_{1,0}.x^{n}_{1,-1}x^{n}_{1,-2}\ldots x^{n}_{1,-Q},\\
x^n_2&=x^{n}_{2,P-1}x^{n}_{2,P-2}\ldots x^{n}_{2,0}.x^{n}_{2,-1}x^{n}_{2,-2}\ldots x^{n}_{2,-Q},\\
     &\vdots\\
x^n_m&=x^{n}_{m,P-1}x^{n}_{m,P-2}\ldots x^{n}_{m,0}.x^{n}_{m,-1}x^{n}_{m,-2}\ldots x^{n}_{m,-Q},
\IEEEstrut	
\end{IEEEeqnarraybox}
\right.
\end{equation*}
where $N=P+Q$.

\item Both IDCS and CBDS deal with 1-D chaotic systems. In contrast, HDDCS extends the system to any finite dimension.

\item Compared to RDCS, IDCS, CBDS and HDDCS need external inputs.

\item The main features of discrete-time RDCS is that all the bits in $x_{n-1}$ will be updated by iteration function $F$ at each iteration operation.
Likewise, all the bits in $x_n$ are updated by iteration function $F$ at each iteration operation. But only one bit of $x_n$ in IDCS is updated by
iteration function $F$ at each iteration operation. CBDS uses multiple random bitwise operations instead of only one in IDCS. HDDCS are similar to
CBDS but can work in higher-dimension.
\end{itemize}
	
\subsection{Network analysis of the state space of HDDCS}

Given a digital chaotic system, a state and its interval is mapped to another one. Considering each state or interval as a node (vertice), and the mapping relation as a directed edge (link), state network of the chaotic system can be build up. As shown in \cite{Hsu:CellMap:IJBC92,Shreim:NetworkCA:2007}, the associated state network can demonstrate some dynamical properties of digital chaotic systems that can not be observed by the previous analytic methods. For the directed graph of $G_F$ in HDDCS, all the possible combinations of $(x_1,x_2,\ldots,x_m)$ are the nodes, and there is a directed edge from node $(\hat{x}_1,\hat{x}_2,\ldots,\hat{x}_m)$ to another node $(\tilde{x}_1,\tilde{x}_2,\ldots,\tilde{x}_m)$ if
\begin{multline*}
(G_F((\hat{s},\hat{u},\ldots,\hat{v}),(\hat{x}_1,\hat{x}_2,\ldots,\hat{x}_m)))_{x_1,x_2,\ldots,x_m}=\\
(\tilde{x}_1,\tilde{x}_2,\ldots,\tilde{x}_m).
\end{multline*} We found that the state network of HDDCS must be strongly connected, namely every node is reachable from
any another one.
\iffalse
The so-called strongly connected network means that a directed edge from node $A$ to node $B$ and from node $B$ to node $A$ exist
simultaneously for any two nodes $A$ and $B$ in the network. As shown in Fig.~\ref{2dexpscc}, it is not a strongly connected network for no path between some pair of nodes of the network.
\fi

In the following, we use 2D-DCS with $N=2$ $(P=2,Q=0)$ to illustrate the property on connectivity.
For example, consider a 2-D digital system uncontrolled by random sequences, as
\begin{equation}
\left\{
\begin{IEEEeqnarraybox}[][c]{l}
\IEEEstrut
	x^n=F_1(x^{n-1},y^{n-1})=\overline{x^{n-1}},\\
	y^n=F_2(x^{n-1},y^{n-1})=\overline{x^{n-1}}\oplus y^{n-1},%\\
	%z^n=x^{n-1}\oplus y^{n-1}\oplus z^{n-1}
\IEEEstrut	
\end{IEEEeqnarraybox} \right.
\label{2dexp}
\end{equation}	
where $\oplus$ denotes the bitwise XOR.

\begin{figure}[!htb]
\centering
\includegraphics[width=0.45\figwidth]{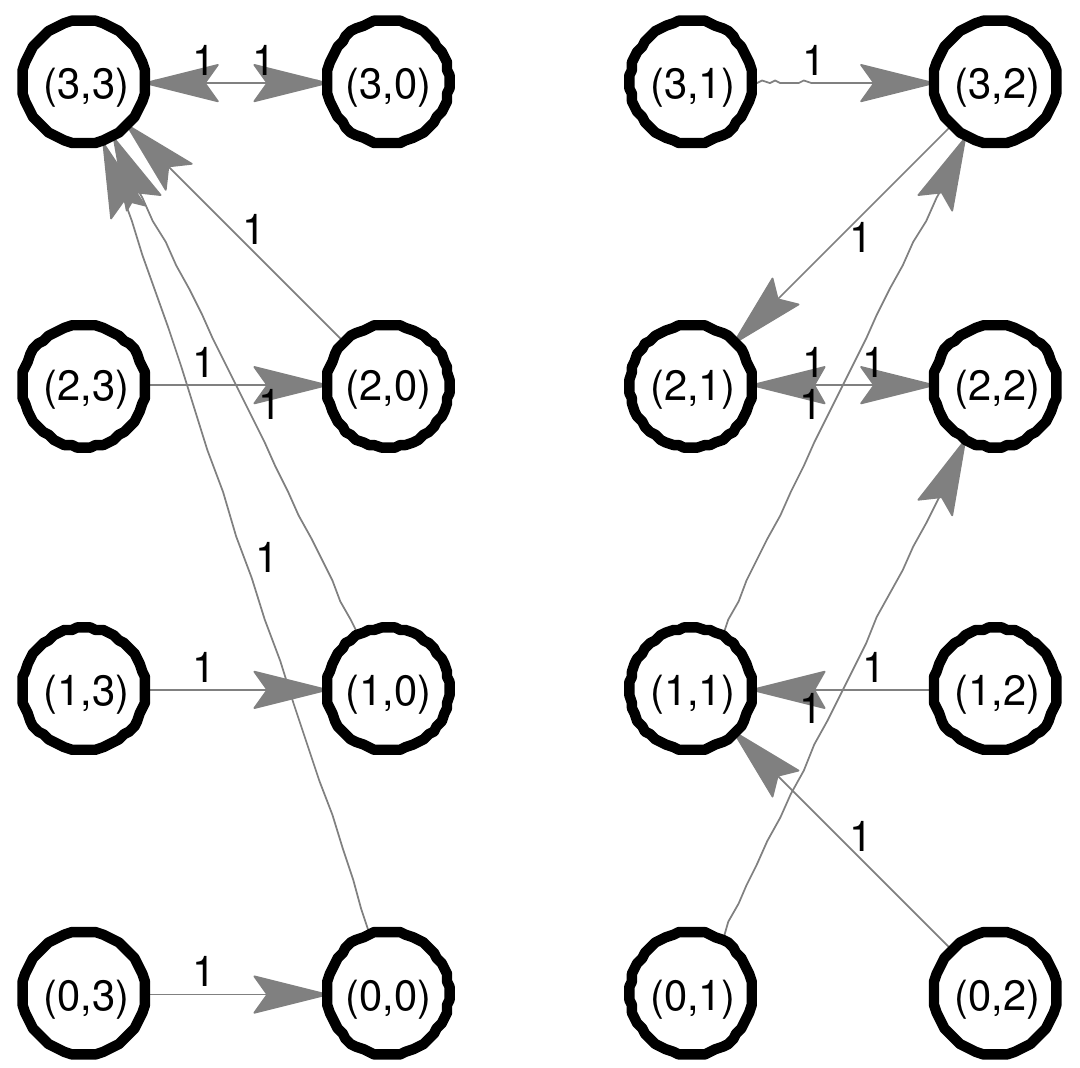}
\caption{The disconnected network for Eq.~(\ref{2dexp}).}
\label{2dexpscc}
\end{figure}

The concrete state transition is shown in Table~\ref{st1}, and the corresponding state transition diagram is shown in Fig.~\ref{2dexpscc}.	
\begin{table}[!t]
\caption{State transition for system~(\ref{2dexp}).}
\label{st1}
\centering
\begin{tabular}{|c|c|c|c|}
\hline
$(x^{n-1},y^{n-1})$ & $(x^{n},y^{n})$ & $(x^{n-1},y^{n-1})$ & $(x^{n},y^{n})$  \\
\hline
$(0,0)$ & $(3,3)$ & $(2,0)$ & $(1,1)$  \\
\hline
$(0,1)$ & $(3,2)$ & $(2,1)$ & $(1,0)$  \\
\hline
$(0,2)$ & $(3,1)$ & $(2,2)$ & $(1,3)$  \\
\hline
$(0,3)$ & $(3,0)$ & $(2,3)$ & $(1,2)$  \\
\hline
$(1,0)$ & $(2,2)$ & $(3,0)$ & $(0,0)$  \\
\hline
$(1,1)$ & $(2,3)$ & $(3,1)$ & $(0,1)$  \\
\hline
$(1,2)$ & $(2,0)$ & $(3,2)$ & $(0,2)$  \\
\hline
$(1,3)$ & $(2,1)$ & $(3,3)$ & $(0,3)$  \\
\hline						
\end{tabular}
\end{table}

\begin{table*}[ht]
\centering
\caption[E]{State Transition of $G_F$ for system~(\ref{2dexp}) in HDDCS.}
\begin{tabular}{c| *{15}{p{4mm}} c}\hline
%\begin{tabular}{c| *{15}{p{4mm}|} c}\hline
\multirow{2}{1.5cm}{$(x^{n-1},y^{n-1})$} &  \multicolumn{16}{c}{$(s^n, u^n)$} \\\cline{2-17}
               & $(0,0)$& $(0,1)$& $(0,2)$& $(0,3)$& $(1,0)$& $(1,1)$& $(1,2)$& $(1,3)$& $(2,0)$& $(2,1)$& $(2,2)$& $(2,3)$& $(3,0)$& $(3,1)$& $(3,2)$& $(3,3)$ \\\hline
$(0,0)$ & $(0,0)$& $(0,1)$& $(0,2)$& $(0,3)$& $(1,0)$& $(1,1)$& $(1,2)$& $(1,3)$& $(2,0)$& $(2,1)$& $(2,2)$& $(2,3)$& $(3,0)$& $(3,1)$& $(3,2)$& $(3,3)$\\\hline
$(0,1)$& $(0,1)$& $(0,0)$& $(0,3)$& $(0,2)$& $(1,1)$& $(1,0)$& $(1,3)$& $(1,2)$& $(2,1)$& $(2,0)$& $(2,3)$& $(2,2)$& $(3,1)$& $(3,0)$& $(3,3)$& $(3,2)$\\\hline
$(0,2)$& $(0,2)$& $(0,3)$& $(0,0)$& $(0,1)$& $(1,2)$& $(1,3)$& $(1,0)$& $(1,1)$& $(2,2)$& $(2,3)$& $(2,0)$& $(2,1)$& $(3,2)$& $(3,3)$& $(3,0)$& $(3,1)$\\\hline
$(0,3)$& $(0,3)$& $(0,2)$& $(0,1)$& $(0,0)$& $(1,3)$& $(1,2)$& $(1,1)$& $(1,0)$& $(2,3)$& $(2,2)$& $(2,1)$& $(2,0)$& $(3,3)$& $(3,2)$& $(3,1)$& $(3,0)$\\\hline
$(1,0)$& $(1,0)$& $(1,0)$& $(1,2)$& $(1,2)$& $(0,0)$& $(0,0)$& $(0,2)$& $(0,2)$& $(3,0)$& $(3,0)$& $(3,2)$& $(3,2)$& $(2,0)$& $(2,0)$& $(2,2)$& $(2,2)$\\\hline
$(1,1)$& $(1,1)$& $(1,1)$& $(1,3)$& $(1,3)$& $(0,1)$& $(0,1)$& $(0,3)$& $(0,3)$& $(3,1)$& $(3,1)$& $(3,3)$& $(3,3)$& $(2,1)$& $(2,1)$& $(2,3)$& $(2,3)$\\\hline
$(1,2)$& $(1,2)$& $(1,2)$& $(1,0)$& $(1,0)$& $(0,2)$& $(0,2)$& $(0,0)$& $(0,0)$& $(3,2)$& $(3,2)$& $(3,0)$& $(3,0)$& $(2,2)$& $(2,2)$& $(2,0)$& $(2,0)$\\\hline
$(1,3)$& $(1,3)$& $(1,3)$& $(1,1)$& $(1,1)$& $(0,3)$& $(0,3)$& $(0,1)$& $(0,1)$& $(3,3)$& $(3,3)$& $(3,1)$& $(3,1)$& $(2,3)$& $(2,3)$& $(2,1)$& $(2,1)$\\\hline
$(2,0)$& $(2,0)$& $(2,1)$& $(2,0)$& $(2,1)$& $(3,0)$& $(3,1)$& $(3,0)$& $(3,1)$& $(0,0)$& $(0,1)$& $(0,0)$& $(0,1)$& $(1,0)$& $(1,1)$& $(1,0)$& $(1,1)$\\\hline
$(2,1)$& $(2,1)$& $(2,0)$& $(2,1)$& $(2,0)$& $(3,1)$& $(3,0)$& $(3,1)$& $(3,0)$& $(0,1)$& $(0,0)$& $(0,1)$& $(0,0)$& $(1,1)$& $(1,0)$& $(1,1)$& $(1,0)$\\\hline
$(2,2)$& $(2,2)$& $(2,3)$& $(2,2)$& $(2,3)$& $(3,2)$& $(3,3)$& $(3,2)$& $(3,3)$& $(0,2)$& $(0,3)$& $(0,2)$& $(0,3)$& $(1,2)$& $(1,3)$& $(1,2)$& $(1,3)$\\\hline
$(2,3)$& $(2,3)$& $(2,2)$& $(2,3)$& $(2,2)$& $(3,3)$& $(3,2)$& $(3,3)$& $(3,2)$& $(0,3)$& $(0,2)$& $(0,3)$& $(0,2)$& $(1,3)$& $(1,2)$& $(1,3)$& $(1,2)$\\\hline
$(3,0)$& $(3,0)$& $(3,0)$& $(3,0)$& $(3,0)$& $(2,0)$& $(2,0)$& $(2,0)$& $(2,0)$& $(1,0)$& $(1,0)$& $(1,0)$& $(1,0)$& $(0,0)$& $(0,0)$& $(0,0)$& $(0,0)$\\\hline
$(3,1)$& $(3,1)$& $(3,1)$& $(3,1)$& $(3,1)$& $(2,1)$& $(2,1)$& $(2,1)$& $(2,1)$& $(1,1)$& $(1,1)$& $(1,1)$& $(1,1)$& $(0,1)$& $(0,1)$& $(0,1)$& $(0,1)$\\\hline
$(3,2)$& $(3,2)$& $(3,2)$& $(3,2)$& $(3,2)$& $(2,2)$& $(2,2)$& $(2,2)$& $(2,2)$& $(1,2)$& $(1,2)$& $(1,2)$& $(1,2)$& $(0,2)$& $(0,2)$& $(0,2)$& $(0,2)$\\\hline
$(3,3)$& $(3,3)$& $(3,3)$& $(3,3)$& $(3,3)$& $(2,3)$& $(2,3)$& $(2,3)$& $(2,3)$& $(1,3)$& $(1,3)$& $(1,3)$& $(1,3)$& $(0,3)$& $(0,3)$& $(0,3)$& $(0,3)$\\
\hline	
\end{tabular}
\label{ch3tbl:vfeatures}
\end{table*}	
	
According to Eq.~(\ref{GF}), Eq.~(\ref{BVN3}), and Eq.~(\ref{2dexp}), one can obtain the iterative form of $G_F(E)_{x,y}$ as
\begin{equation} \left\{
\begin{IEEEeqnarraybox}[][c]{l}
\IEEEstrut
x^n=x^{n-1}\cdot\overline{s^{n}}+(\overline{x^{n-1}}\cdot s^n),\\
y^n=y^{n-1}\cdot\overline{u^{n}}+((\overline{x^{n-1}}\oplus y^{n-1})\cdot u^n),
%z^n=x^{n-1}\oplus y^{n-1}\oplus z^{n-1}
\IEEEstrut	
\end{IEEEeqnarraybox}\right.
\label{gfxy}
\end{equation}
where $s=s^1s^2s^3\ldots$ and $u=u^1u^2u^3\ldots$ are two random sequences generated by TRNG.

\begin{figure}[!htb]
\centering
\includegraphics[width=0.8\figwidth]{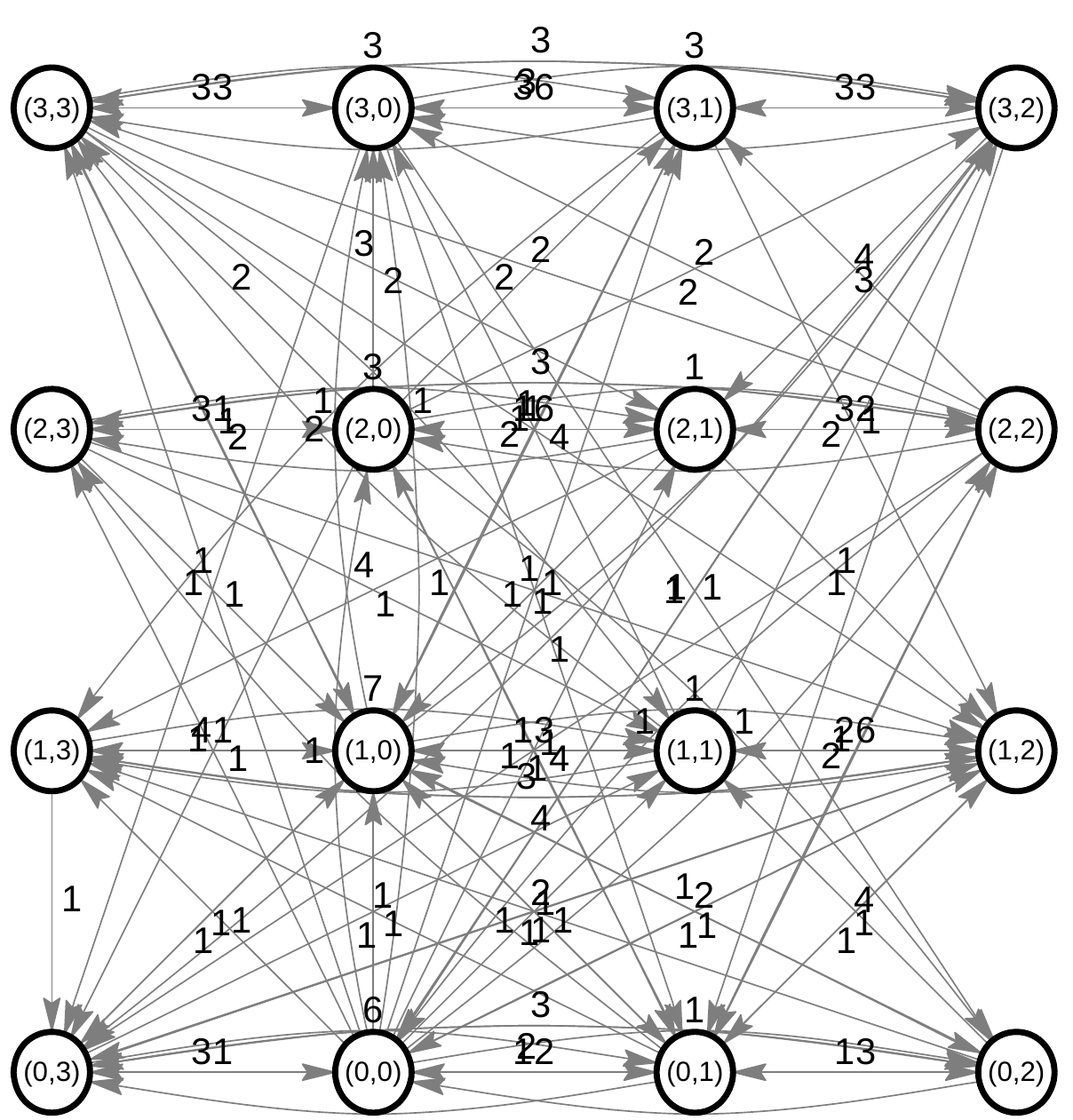}
\caption{The strongly connected network for $G_F$ of system~(\ref{2dexp}).}
\label{2dddcsexpscc}
\end{figure}

According to Eq.~(\ref{gfxy}), the concrete state transition of $G_F(E)_{x,y}$ is shown in Table~\ref{ch3tbl:vfeatures}, and the corresponding state network is shown in Fig.~\ref{2dddcsexpscc} , which is a strongly connected network. In the next section, we will prove that HDDCS satisfies the Devaney's definition of chaos if the state network of its $G_F$ is strongly connected.
	
\section{Chaotic performance of HDDCS}
\label{ChaosProofHDDCS}

In this section, we prove that the map of HDDCS satisfies Devaney's definition of chaos, \textit{i.e.},
its periodic points are dense in its definitional domain and it is transitive \cite{Banks:definition:AMM1992}.

\subsection{Dense periodic points of HDDCS}

To prove the existence of a dense set of periodic points and the transitivity property of the discrete dynamical system defined previously,
the following lemma is firstly proven.
\begin{lemma}
\label{Lemma1}
Let $w\in\{s, u, \cdots, v\}$, $w=w^1w^2w^3 \ldots w^n \ldots$ and $\hat{w} = \hat{w}^1\hat{w}^2\hat{w}^3 \ldots \hat{w}^n \ldots$,
the metric distance $d$ satisfies that if
$w^i = \hat{w}^i$ for $i=1,2,3,\ldots n$, then
\begin{equation*}
d(w,\hat{w}) \leq \frac{1}{2^{Nn}},
\end{equation*}
where $w^k, \hat{w}^k \in [0,2^P-2^{-Q}]$ for $k\in \mathbb{Z}^+$.
\end{lemma}
\begin{IEEEproof}
If $w^i = \hat{w}^i$ for $i = 1, 2, \ldots ,n$ , then
\begin{IEEEeqnarray*}{rCl}
d(w,\hat{w}) & =    & \sum_{i=1}^{n}\frac{|w^i- \hat{w}^i|}{2^{Ni}} + \sum_{i=n+1}^{\infty}\frac{|w^i - \hat{w}^i|}{2^{Ni}} \\
			 & =    & \sum_{i=n+1}^{\infty}\frac{|w^i - \hat{w}^i|}{2^{Ni}} \\
             & \leq & \sum_{i=n+1}^{\infty}\frac{2^P-2^{-Q}}{2^{Ni}}\\
%			 & \leq & (2^P-2^{-Q}) \times \frac{\frac{1}{2^{Nn+N}}}{1-\frac{1}{2^N}} \\
             & =    &  \frac{2^P-2^{-Q}}{2^{N}-1}\cdot \frac{1}{2^{Nn}} \leq \frac{1}{2^{Nn}}.
\end{IEEEeqnarray*}
Due to the definition of the proposed distance: for any $m\leqslant n$, if $w^m=\hat{w}^m$, then $d(w,\hat{w}) \leq \frac{1}{2^{Nn}}$.
%Due to the definition of the proposed distance: for any $m\leqslant n$, if $w^m\neq\hat{w}^m$, then $d(w,\hat{w})\geqslant \frac{1}{2^{Nn}}$. The contraposition is the desired result: if $d(w,\hat{w})\leqslant \frac{1}{2^{Nn}}$, then $w^i=\hat{w}^i(i=1,2,\ldots,n)$.
\end{IEEEproof}
The lemma can let us quickly determine whether the two sequences are close to each other.
From intuitive observation, we can assure two sequences are close to each other as long as they have a considerable number of consistent foregoing entries.
\begin{theorem}
The periodic points of HDDCS are dense in the metric space $(\mathcal{E}, d)$.
\end{theorem}
\begin{IEEEproof}
For any given $\varepsilon \in (0, 1)$ and a periodic point
$((\tilde{s},\tilde{u},\ldots,\tilde{v})(\tilde{x}_1,\tilde{x}_2,\ldots,\tilde{x}_m))\in \mathcal{E}$, let us try to prove that the dense periodic points in $(\mathcal{E}, d)$ can always be found in the neighborhood of distance $\varepsilon$ of any point
$((\hat{s},\hat{u},\ldots,\hat{v}),(\hat{x}_1,\hat{x}_2,\ldots,\hat{x}_m)) \in \mathcal{E}$ as
\begin{multline*}
\label{period1}
d(((\tilde{s},\tilde{u},\ldots,\tilde{v}),(\tilde{x}_1,\tilde{x}_2,\ldots,\tilde{x}_m)),\\
((\hat{s},\hat{u},\ldots,\hat{v}),(\hat{x}_1,\hat{x}_2,\ldots,\hat{x}_m)))<\varepsilon.
\end{multline*}
Without loss of generality, we assume that the general form of $((\hat{s},\hat{u},\ldots,\hat{v}),(\hat{x}_1,\hat{x}_2,\ldots,\hat{x}_m)))$
is
\begin{multline*}
((\hat{s},\hat{u},\ldots,\hat{v}),(\hat{x}_1,\hat{x}_2,
\ldots,\hat{x}_m)))\\
=(((\hat{s}^1\hat{s}^2\ldots \hat{s}^{k_0}\ldots \hat{s}^n\ldots),(\hat{u}^1\hat{u}^2\ldots \hat{u}^{k_0}\ldots \hat{u}^n\ldots),\\
\qquad  \ldots,(\hat{v}^1\hat{v}^2\ldots \hat{v}^{k_0}\ldots \hat{v}^n\ldots)),(\hat{x}_1,\hat{x}_2,
\ldots,\hat{x}_m))
\in \mathcal{E}.
\end{multline*}

Given $\varepsilon<2^{-Q}$, if point $(\hat{x}_1,\hat{x}_2,\ldots,\hat{x}_m)$ and $(\tilde{x}_1,\tilde{x}_2,\ldots,\tilde{x}_m)$ do not coincide in
the $m$-dimensional space, we can obtain $(\tilde{x}_1,\tilde{x}_2,\ldots,\tilde{x}_m)\neq(\hat{x}_1,\hat{x}_2,\ldots,\hat{x}_m)$ such that
\begin{equation*}
\begin{IEEEeqnarraybox}[\IEEEeqnarraystrutmode\IEEEeqnarraystrutsizeadd{2pt}{2pt}][c]{rCl}
\IEEEeqnarraymulticol{3}{l}{d_{x}((\tilde{x}_1,\tilde{x}_2,\ldots,\tilde{x}_m),(\hat{x}_1,\hat{x}_2,\ldots,\hat{x}_m)}\\
\quad   &=    &\sqrt{(\tilde{x}_1-\hat{x}_1)^2+(\tilde{x}_2-\hat{x}_2)^2+\ldots+(\tilde{x}_m-\hat{x}_m)^2}\\
\quad  	&\geq & 2^{-Q}.
\end{IEEEeqnarraybox}
\end{equation*}
Then, one has
\begin{multline*}
d(((\hat{s},\hat{u},\ldots,\hat{v}),(\hat{x}_1,\hat{x}_2,\ldots,\hat{x}_m)),\\((\tilde{s},\tilde{u},\ldots,\tilde{v}),(\tilde{x}_1,\tilde{x}_2,
\ldots,\tilde{x}_m)))\geq\varepsilon.
\end{multline*}
So, to satisfy 		
\begin{multline*}
d(((\hat{s},\hat{u},\ldots,\hat{v}),(\hat{x}_1,\hat{x}_2,\ldots,\hat{x}_m)),\\((\tilde{s},\tilde{u},\ldots,\tilde{v}),(\tilde{x}_1,\tilde{x}_2,
\ldots,\tilde{x}_m)))<\varepsilon,		
\end{multline*}
we must first set $\tilde{x}_1=\hat{x}_1,\tilde{x}_2=\hat{x}_2,\ldots, \tilde{x}_m=\hat{x}_m$. In case
\begin{multline*}
d(((\hat{s},\hat{u},\ldots,\hat{v}),(\hat{x}_1,\hat{x}_2,\ldots,\hat{x}_m)),\\((\tilde{s},\tilde{u},\ldots,\tilde{v}),(\tilde{x}_1,\tilde{x}_2,
\ldots,\tilde{x}_m)))<\varepsilon,		
\end{multline*}
we should consider to prove  that $((\tilde{s},\tilde{u},\ldots,\tilde{v}),(\tilde{x}_1,\tilde{x}_2,\ldots,\tilde{x}_m))$ is a periodic point.
If the first $k_0$ elements of $\hat{s}$ and $\tilde{s}$ are the same, then $d_s(\hat{s},\tilde{s})<2^
{-Nk_0}<\varepsilon$. Referring to Lemma~\ref{Lemma1}, similar results can be obtained as $k_0$ elements of $\hat{u}, \ldots, \hat{v}$ and $\tilde{u}, \ldots, \hat{v}$
are the same and
\begin{equation*}
d_u(\hat{u},\tilde{u})<2^
{-Nk_0}<\varepsilon,\ldots,d_v(\hat{v},\tilde{v})<2^
{-Nk_0}<\varepsilon.
\end{equation*}
So, $\forall\ \varepsilon<1$, an integer $k_0$ satisfying the relation $d_s(\hat{s},\tilde{s})+d_u(\hat{u},\tilde{u})+\ldots+d_v(\hat{v},\tilde{v})<m\times2^
{-Nk_0}<\varepsilon$ can always be found. For instance, to make $m\times2^
{-Nk_0}<\varepsilon$ hold, the value of $k_0$ can be set as
\begin{equation*}
k_0=\lfloor(\log_2m-\log_2\varepsilon)/N \rfloor+1.
\end{equation*}
After $k_0$-th iteration, one has
\begin{multline*}
(G_F^{k_0}((\tilde{s},\tilde{u},\ldots,\tilde{v}),(\hat{x}_1,\hat{x}_2,\ldots,\hat{x}_m)))_{x_1,x_2,\ldots,x_m}\\
=(\hat{x}_1,\hat{x}_2,\ldots,\hat{x}_m).
\end{multline*}

The above equation shows that HDDCS starts from $(\hat{x}_1, \hat{x}_2, \ldots, \hat{x}_m)$, then returns back to it after $k_0$ iterations.
This means that a periodic point $(\hat{x}_1,\hat{x}_2,\ldots,\hat{x}_m)$ is found out:
\begin{equation*}
\begin{IEEEeqnarraybox}[\IEEEeqnarraystrutmode\IEEEeqnarraystrutsizeadd{2pt}{2pt}][c]{rCl}
\IEEEeqnarraymulticol{3}{l}{((\tilde{s},\tilde{u},\ldots,\tilde{v}),(\hat{x}_1,\hat{x}_2,\ldots,\hat{x}_m))}\\
		& = & (((s^1 s^2\ldots s^{k_0}s^1s^2\ldots s^{k_0}\ldots), (u^1 u^2\ldots u^{k_0}u^1u^2\ldots u^{k_0}\ldots),\\
        &   & \ldots, (v^1v^2\ldots v^{k_0}v^1v^2\ldots v^{k_0}\ldots)), (\hat{x}_1,\hat{x}_2,\ldots,\hat{x}_m)),
\end{IEEEeqnarraybox}
\end{equation*} which satisfies
\begin{multline*}
G_F^{k_0}((\tilde{s},\tilde{u},\ldots,\tilde{v}),(\hat{x}_1,\hat{x}_2,\ldots,\hat{x}_m))\\
=((\tilde{s},\tilde{u},\ldots,\tilde{v}),(\hat{x}_1,\hat{x}_2,\ldots,\hat{x}_m)),
\end{multline*}
and
\begin{equation*}
\begin{IEEEeqnarraybox}[\IEEEeqnarraystrutmode\IEEEeqnarraystrutsizeadd{2pt}{2pt}][c]{rCl}
\IEEEeqnarraymulticol{3}{l}{d(((\hat{s},\hat{u},\ldots,\hat{v}),(\hat{x}_1,\hat{x}_2,\ldots,\hat{x}_m)),}\\
 &   & ((\tilde{s},\tilde{u},\ldots,\tilde{v}),(\tilde{x}_1,\tilde{x}_2,\ldots,\tilde{x}_m))) \\
 &= & d_s(\hat{s},\tilde{s})+d_u(\hat{u},\tilde{u})+\ldots+d_v(\hat{v},\tilde{v})\\
 &  &  +d_{x}((\hat{x}_1,\hat{x}_2,\ldots,\hat{x}_m),(\hat{x}_1,\hat{x}_2,\ldots,\hat{x}_m))\\
 &= & d_s(\hat{s},\tilde{s})+d_u(\hat{u},\tilde{u})+\ldots+d_v(\hat{v},\tilde{v})\\
 &< & \varepsilon.
\end{IEEEeqnarraybox}
\end{equation*}

If after $k_0$-th iteration, one get
\begin{multline*}			
(G_F^{k_0}((\tilde{s},\tilde{u},\ldots,\tilde{v}),(\hat{x}_1,\hat{x}_2,\ldots,\hat{x}_m)))_{x_1,x_2,\ldots,x_m}\\
\neq(\hat{x}_1,\hat{x}_2,\ldots,\hat{x}_m)		
\end{multline*}
 and 		
\begin{multline*}	
(G_F^{k_0}((\tilde{s},\tilde{u},\ldots,\tilde{v}),(\hat{x}_1,\hat{x}_2,\ldots,\hat{x}_m)))_{x_1,x_2,\ldots,x_m}\\
=(x_1',x_2',\ldots,x_m').
\end{multline*}	
	
Because $G_F$ is strongly connected, there is at least one path from the state $(x_1',x_2',\ldots,x_m')$ to the
state $(\hat{x}_1,\hat{x}_2,\ldots,\hat{x}_m)$, after another iteration of $i_0$ times,
where $i_0$ is equal to the number of edges in the connected path between the state $(x_1',x_2',\ldots,x_m')$ and the state
$(\hat{x}_1,\hat{x}_2,\ldots,\hat{x}_m)$. By making the equation
\begin{multline*}
(G_F^{k_0+i_0}((\tilde{s},\tilde{u},\ldots,\tilde{v}),(\hat{x}_1,\hat{x}_2,\ldots,\hat{x}_m)))_{x_1,x_2,\ldots,x_m}\\
=(\hat{x}_1,\hat{x}_2,\ldots,\hat{x}_m)
\end{multline*}
hold, a periodic point is found by checking
\begin{equation*}
\begin{IEEEeqnarraybox}[\IEEEeqnarraystrutmode\IEEEeqnarraystrutsizeadd{2pt}{2pt}][c]{rCl}
\IEEEeqnarraymulticol{3}{l}{((\tilde{s},\tilde{u},\ldots,\tilde{v}),(\hat{x}_1,\hat{x}_2,\ldots,\hat{x}_m))}\\
&= &(((s^1 s^2\ldots s^{k_0}s^{k_0+1}s^{k_0+2}\ldots s^{k_0+i_0}s^1s^2\ldots s^{k_0}\\
&  & s^{k_0+1}s^{k_0+2}\ldots s^{k_0+i_0}\ldots),(u^1 u^2\ldots u^{k_0}u^{k_0+1}\\
&  & u^{k_0+2}\ldots u^{k_0+i_0}u^1u^2\ldots u^{k_0}u^{k_0+1}u^{k_0+2}\ldots\\
&  & u^{k_0+i_0}\ldots),\ldots,(v^1 v^2\ldots v^{k_0}v^{k_0+1}v^{k_0+2}\ldots\\
&  & v^{k_0+i_0}v^1v^2\ldots v^{k_0}v^{k_0+1}v^{k_0+2}\ldots v^{k_0+i_0}\ldots))\\
&  & (\hat{x}_1,\hat{x}_2,\ldots,\hat{x}_m)),
\end{IEEEeqnarraybox}
\end{equation*}
which satisfies
\begin{equation*}
\begin{IEEEeqnarraybox}[\IEEEeqnarraystrutmode\IEEEeqnarraystrutsizeadd{2pt}{2pt}][c]{rCl}
\IEEEeqnarraymulticol{3}{l}{d(((\hat{s},\hat{u},\ldots,\hat{v}),(\hat{x}_1,\hat{x}_2,\ldots,\hat{x}_m))),}\\
& & \quad\quad\quad((\tilde{s},\tilde{u},\ldots,\tilde{v}),(\tilde{x}_1,\tilde{x}_2,\ldots,\tilde{x}_m)))\\	
&=&d_s(\hat{s},\tilde{s})+d_u(\hat{u},\tilde{u})+\ldots+d_v(\hat{v},\tilde{v})\\
& &\ +d_{x}((\hat{x}_1,\hat{x}_2,\ldots,\hat{x}_m),(\hat{x}_1,\hat{x}_2,\ldots,\hat{x}_m))\\
&=&d_s(\hat{s},\tilde{s})+d_u(\hat{u},\tilde{u})+\ldots+d_v(\hat{v},\tilde{v})\\
& <&\varepsilon.
\end{IEEEeqnarraybox}
\end{equation*}

\begin{figure}[!htb]
\centering
\includegraphics[width=\figwidth]{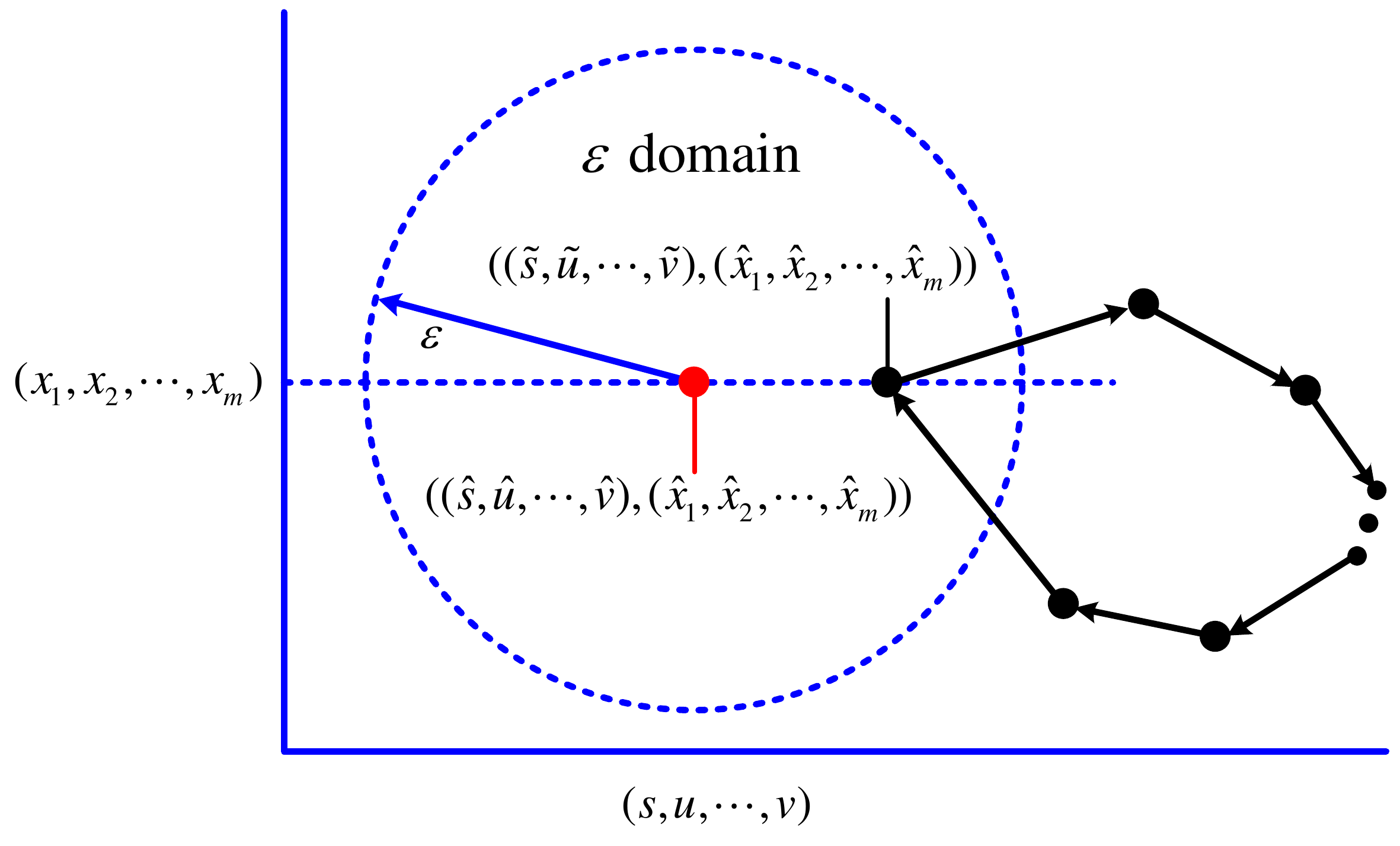}
\caption{The diagram of the periodic points of HDDCS.}
\label{Dense_periodic_points}
\end{figure}

In summary, the periodic points of $G_F$ are dense in metric space $(\mathcal{E},d)$, as shown in Fig.~\ref{Dense_periodic_points}.
\end{IEEEproof}

\subsection{Transitive property of HDDCS}

\begin{theorem}
Function $G_F$ is topological transitive in the metric space $(\mathcal{E},d)$.
\end{theorem}
\begin{IEEEproof}
The so-called topological transitivity of function $G_F$ in metric space $(G_F,\mathcal{E})$
means that there always exists $n_0>0$ satisfying $G_F^{n_0}(U')\cap U''\neq \varnothing$ for any nonempty open sets $U'$ and $U''$.
%which has a point $((s',u',\ldots,v'),(x_1',x_2',\ldots,x_m'))$ as the center and $r'$ as the sphere radius, and
%$U''$ which has a point $((s'',u'',\ldots,v''),(x_1'',x_2'',\ldots,x_m''))$ as the center and $r''$ as the sphere radius.
Let $((s',u',\ldots,v'),(x_1',x_2',\ldots,x_m'))\in U'\subseteq \mathcal{E}$ denote the center of $U'$, which can be represented as
\begin{equation*}
\begin{IEEEeqnarraybox}[\IEEEeqnarraystrutmode\IEEEeqnarraystrutsizeadd{2pt}{2pt}][c]{rCl}
\IEEEeqnarraymulticol{3}{l}{
((s',u',\ldots,v'),(x_1',x_2',\ldots,x_m'))}\\
\quad & = & (((s'^1s'^2\ldots s'^{n_0}\ldots s'^n\ldots ),\\
\quad &   & (u'^1u'^2\ldots u'^{n_0}\ldots u'^n\ldots ),\ldots,\\
\quad &   & (v'^1v'^2\ldots v'^{n_0}\ldots v'^n\ldots )),\\
\quad &   & (x_1',x_2',\ldots,x_m').
\end{IEEEeqnarraybox}
\end{equation*}
The center of $U''$, $((s'',u'',\ldots,v''),(x_1'',x_2'',\ldots,x_m''))$ can also be presented as
\begin{equation*}
\begin{IEEEeqnarraybox}[\IEEEeqnarraystrutmode\IEEEeqnarraystrutsizeadd{2pt}{2pt}][c]{rCl}
\IEEEeqnarraymulticol{3}{l}{
((s'',u'',\ldots,v''),(x_1'',x_2'',\ldots,x_m''))}\\
&=&(((s''^1s''^2\ldots s''^{n_0}\ldots s''^n\ldots ),\\
& &u''^1u''^2\ldots u''^{n_0}\ldots u''^n\ldots ),\ldots,\\
& &(v''^1v''^2\ldots v''^{n_0}\ldots v''^n\ldots )),\\
& &x_1'',x_2'',\ldots,x_m''),
\end{IEEEeqnarraybox}
\end{equation*}
and a point in $U'$ is denoted as $((\tilde{s},\tilde{u},\ldots,\tilde{v}),(\tilde{x}_1,\tilde{x}_2,\ldots,\tilde{x}_m)\in U'\subseteq \mathcal{E}$.

If the sphere radius of $U'$, $r'$, is less than $2^{-Q}$ and points $(x_1',x_2',\ldots,x_m')$ and $(\tilde{x}_1,\tilde{x}_2,\ldots,\tilde{x}_m)$ do not coincide in the $m$-dimensional space,
%For any $r'<2^{-Q}$, if points $(x_1',x_2',\ldots,x_m')$ and $(\tilde{x}_1,\tilde{x}_2,\ldots,\tilde{x}_m)$ do not coincide in the $m$-dimensional space,
one has
\begin{equation*}
((\tilde{s},\tilde{u},\ldots,\tilde{v}),(\tilde{x}_1,\tilde{x}_2,\ldots,\tilde{x}_m)\notin U'
\end{equation*}
as
\begin{equation*}
\begin{IEEEeqnarraybox}[\IEEEeqnarraystrutmode\IEEEeqnarraystrutsizeadd{2pt}{2pt}][c]{rCl}
\IEEEeqnarraymulticol{3}{l}{
d_{x}((\tilde{x}_1,\tilde{x}_2,\ldots,\tilde{x}_m),(x_1',x_2',\ldots,x_m'))}\\
\quad & =   &\sqrt{(\tilde{x}_1-x'_1)^2+(\tilde{x}_2-x'_2)^2+\ldots+(\tilde{x}_m-x'_m)^2}\\
      &\geq & 2^{-Q}>r',
\end{IEEEeqnarraybox}
\end{equation*}
and $(\tilde{x}_1,\tilde{x}_2,\ldots,\tilde{x}_m)\neq(x_1',x_2',\ldots,x_m')$. So, to satisfy
\begin{equation*}
((\tilde{s},\tilde{u},\ldots,\tilde{v}),(\tilde{x}_1,\tilde{x}_2,\ldots,\tilde{x}_m)\in U',
\end{equation*}
we must set
\begin{equation*}
(\tilde{x}_1,\tilde{x}_2,\ldots,\tilde{x}_m)=(x_1',x_2',\ldots,x_m')
\end{equation*}
to obtain
\begin{equation*}
d_{x}((\tilde{x}_1,\tilde{x}_2,\ldots,\tilde{x}_m),(x_1',x_2',\ldots,x_m'))=0.
\end{equation*}

If the first $k_0$ elements of $s'$ and $\tilde{s}$ are the same, then $d_s(s',\tilde{s})<2^
{-Nk_0}$. From Lemma~\ref{Lemma1}, the similar results can be obtained when the first $k_0$ elements of $u', \ldots, v'$ and $\tilde{u}, \ldots,\tilde{v}$ are the same and
\begin{equation*}
d_u(u',\tilde{u})<2^
{-Nk_0}<\varepsilon,\ldots,d_v(v',\tilde{v})<2^
{-Nk_0}<\varepsilon.
\end{equation*}
So, $\forall\ r'<1$, an integer $k_0$ satisfying $d_s(s',\tilde{s})+d_u(u',\tilde{u})+\ldots+d_v(v',\tilde{v})<m\times2^
{-Nk_0}< r'$ can always be found. For instance, to satisfy $m\times2^
{-Nk_0}< r'$, the value of $k_0$ can set as
\begin{equation*}
k_0=\lfloor(\log_2m-\log_2 r')/N\rfloor+1.
\end{equation*}
If after the $k_0$-th iteration, equality
\begin{multline*}
(G_F^{k_0}((\tilde{s},\tilde{u},\ldots,\tilde{v}),(\tilde{x}_1,\tilde{x}_2,\ldots,\tilde{x}_m)))_{x_1,x_2,\ldots,x_m}\\
=(x_1'',x_2'',\ldots,x_m'')
\end{multline*}
exists, then $n_0=k_0$ is found, and
\begin{equation*}
\begin{IEEEeqnarraybox}[\IEEEeqnarraystrutmode\IEEEeqnarraystrutsizeadd{2pt}{2pt}][c]{rCl}
\IEEEeqnarraymulticol{3}{l}{
((\tilde{s},\tilde{u},\ldots,\tilde{v}),(\tilde{x}_1,\tilde{x}_2,\ldots,\tilde{x}_m))}\\	
& =   & (((s'^1 s'^2\ldots s'^{n_0}s''^1s''^2\ldots s''^{n}\ldots),(u'^1 u'^2\ldots u'^{n_0}\\
&     & u''^1u''^2\ldots u''^{n}\ldots),\ldots,(v'^1 v'^2\ldots v'^{n_0}v''^1v''^2\ldots\\
&     & v''^{n}\ldots)),(x_1',x_2',\ldots, x_m'))\in U',
\end{IEEEeqnarraybox}
\end{equation*}
which satisfies
\begin{equation*}
\begin{IEEEeqnarraybox}[\IEEEeqnarraystrutmode\IEEEeqnarraystrutsizeadd{2pt}{2pt}][c]{rCl}
\IEEEeqnarraymulticol{3}{l}{G_F^{n_0}((\tilde{s},\tilde{u},\ldots,\tilde{v}),(\tilde{x}_1,\tilde{x}_2,\ldots,\tilde{x}_m))}\\
& =  & ((s'',u'',\ldots,v''),(x_1'',x_2'',\ldots,x_m''))\\
&\in & G_F^{n_0}(U')\cap U''
\end{IEEEeqnarraybox}.
\end{equation*}
So,
\begin{equation*}
G_F^{n_0}(U')\cap U''\neq \varnothing
\end{equation*}
hold.

If after the $k_0$-th iteration, inequality
\begin{multline*}
(G_F^{k_0}((\tilde{s},\tilde{u},\ldots,\tilde{v}),(\tilde{x}_1,\tilde{x}_2,\ldots,\tilde{x}_m)))_{x_1,x_2,\ldots,x_m}\\
\neq(x_1'',x_2'',\ldots,x_m''),
\end{multline*}
holds, set $(G_F^{k_0}((\tilde{s},\tilde{u},\ldots,\tilde{v}),(\tilde{x}_1,\tilde{x}_2,\ldots,\tilde{x}_m)))_{x_1,x_2,\ldots,x_m}=(x_1''',x_2''',\ldots,x_m''').$
Because $G_F$ is strongly connected, there is at least one path from the state $(x_1''',x_2''',\ldots,x_m''')$ to the state $(x_1'',x_2'',\ldots,x_m'')$ after
other $i_0$ iterations. As
\begin{multline*}
G_F^{i_0}((\tilde{s},\tilde{u},\ldots,\tilde{v}),(x_1''',x_2''',\ldots,x_m'''))\\
=((s'',u'',\ldots,v''),(x_1'',x_2'',\ldots,x_m''))
\end{multline*}
hold, $n_0=k_o+i_0$ is found, and
\begin{equation*}
\begin{IEEEeqnarraybox}[\IEEEeqnarraystrutmode\IEEEeqnarraystrutsizeadd{2pt}{2pt}][c]{rCl}
\IEEEeqnarraymulticol{3}{l}{((\tilde{s},\tilde{u},\ldots,\tilde{v}),(\tilde{x}_1,\tilde{x}_2,\ldots,\tilde{x}_m))}\\
& = & (((s'^1 s'^2\ldots s'^{k_0}s^{k_0+1}s^{k_0+2}\ldots s^{k_0+i_0}s''^1s''^2\ldots \\
&   & s''^{n}\ldots),(u'^1 u'^2\ldots u'^{k_0}u^{k_0+1}u^{k_0+2}\ldots u^{k_0+i_0}\\
&   & u''^1u''^2\ldots u''^{n}\ldots),\ldots,(v'^1 v'^2\ldots v'^{k_0}v^{k_0+1}\\
&   & v^{k_0+2}\ldots v^{k_0+i_0}v''^1v''^2\ldots v''^{n}\ldots)),\\
&&(x_1',x_2',\ldots, x_m'))\\
& \in & U',
\end{IEEEeqnarraybox},=
\end{equation*}			
which satisfies
\begin{equation*}
\begin{IEEEeqnarraybox}[\IEEEeqnarraystrutmode\IEEEeqnarraystrutsizeadd{2pt}{2pt}][c]{rCl}
\IEEEeqnarraymulticol{3}{l}{
G_F^{n_0}((\tilde{s},\tilde{u},\ldots,\tilde{v}),(\tilde{x}_1,\tilde{x}_2,\ldots,\tilde{x}_m))}\\
& =   &((s'',u'',\ldots,v''),(x_1'',x_2'',\ldots,x_m''))\\
& \in & G_F^{n_0}(U')\cap U''.
\end{IEEEeqnarraybox}
\end{equation*}
So, one has
\begin{equation*}
G_F^{n_0}(U')\cap U''\neq \varnothing.
\end{equation*}
In summary, $G_F$ is transitive in the metric space $(\mathcal{E},d)$.
\end{IEEEproof}
If a dynamical system on a metric space is transitive and has dense periodic points, then it has sensitive dependence on initial conditions
\cite{Banks:definition:AMM1992}. In other words, if the state network of HDDCS is strongly connected, we can prove that it is chaotic in the sense of Devaney's
definition of chaos.
	
\section{Lyapunov exponents of a class of HDDCS}
\label{Lyapunovexponent}

In this section, the Lyapunov exponents of HDDCS with $N=P$ $(Q=0)$ are estimated.

\subsection{The general expression of equivalent decimal for $G_F$}

Set the binary form of $m$-dimensional array of $N$-bit integers as
\begin{equation*}
\left\{
\begin{IEEEeqnarraybox}[][c]{ll}
\IEEEstrut
x_1 & =x_{1,N-1}x_{1,N-2}\ldots x_{1,0},\\
x_2 & =x_{2,N-1}x_{2,N-2}\ldots x_{2,0},\\
    & \vdots\\
x_m & =x_{m,N-1}x_{m,N-2}\ldots x_{m,0},
\IEEEstrut	
\end{IEEEeqnarraybox}
\right.
\end{equation*}
where $x_{i,j}\in\{0,1\}$ $(i=1,2,\ldots,m; j=N-1,N-2,\ldots,0)$.

The general form of corresponding decimal integer is  obtained as
\begin{equation}
\left\{
\begin{IEEEeqnarraybox}[][c]{ll}
\IEEEstrut
X_1 & =\sum_{k=0}^{N-1}(x_{1,k}\cdot 2^k).\\
X_2 & =\sum_{k=0}^{N-1}(x_{2,k}\cdot 2^k).\\
    & \vdots\\
X_m &=\sum_{k=0}^{N-1}(x_{m,k}\cdot 2^k).
\IEEEstrut	
\end{IEEEeqnarraybox} \right.
\label{decimal_integer}
\end{equation}

Random number of $m$ sequences is expressed in corresponding decimal fraction as
\begin{equation}
\left\{
\begin{IEEEeqnarraybox}[][c]{ll}
\IEEEstrut
S&=\sum_{k=1}^{+\infty}(s^k\cdot 2^{-Nk}),\\
U&=\sum_{k=1}^{+\infty}(u^k\cdot 2^{-Nk}),\\
 & \vdots\\
V&=\sum_{k=1}^{+\infty}(v^k\cdot 2^{-Nk}),
\IEEEstrut	
\end{IEEEeqnarraybox}
\right.
\label{decimal_fraction}
\end{equation}
where $2^{-Nk}(k=1,2,\ldots)$ denote weights.

According to Eq.~(\ref{decimal_integer}) and (\ref{decimal_fraction}) added together, the general form of the corresponding decimal number is obtained as
\begin{equation*}
\left\{
\begin{IEEEeqnarraybox}[][c]{ll}
\IEEEstrut
y_1&=\sum_{k=0}^{N-1}(x_{1,k}\cdot 2^k)+\sum_{k=1}^{+\infty}(s^k\cdot 2^{-Nk}),\\
y_2&=\sum_{k=0}^{N-1}(x_{2,k}\cdot 2^k)+\sum_{k=1}^{+\infty}(u^k\cdot 2^{-Nk}),\\
   &\vdots\\
y_m&=\sum_{k=0}^{N-1}(x_{m,k}\cdot 2^k)+\sum_{k=1}^{+\infty}(v^k\cdot 2^{-Nk}).
\IEEEstrut	
\end{IEEEeqnarraybox}
\right.
\end{equation*}

According to the chaos generation strategy controlled by random sequences, the general form of $m$-Dimensional digital discrete-time iterative equation
can be presented as
\begin{equation}
\left\{
\begin{IEEEeqnarraybox}[][c]{lll}
\IEEEstrut
g_1(X)&=&(x_1\cdot\overline{s^1})+(F_1(\cdot)\cdot s^1),\\
g_2(X)&=&(x_2\cdot\overline{u^1})+(F_2(\cdot)\cdot u^1),\\
&\vdots&\\
g_m(X)&=&(x_m\cdot\overline{v^1})+(F_m(\cdot)\cdot v^1),
\IEEEstrut
\end{IEEEeqnarraybox} \right.
\label{decimal_integer2}
\end{equation}
where $X=(X_1,X_2,\ldots,X_m)$ and
\begin{equation*}
\left\{
\begin{IEEEeqnarraybox}[][c]{lll}
\IEEEstrut
F_1(\cdot)&\triangleq & F_1(x_{1},x_{2},\ldots, x_{m}),\\
F_2(\cdot)&\triangleq & F_2(x_{1},x_{2},\ldots, x_{m}),\\
          & \vdots    &\\
F_m(\cdot)&\triangleq &F_m(x_{1},x_{2},\ldots, x_{m}).
\IEEEstrut
\end{IEEEeqnarraybox} \right.
\end{equation*}

Separately shifts one value in each one-sided infinite sequence ($s=s^1s^2\ldots s^n \ldots$, $u=u^1u^2\ldots u^n \ldots$,
$\ldots$, $v=v^1v^2\ldots v^n \ldots$), the first value turn into $s^2$, $u^2$, $\ldots$, $v^2$ individually,
the corresponding weight is $2^{-N}$. So, general form of the corresponding decimal fraction after shifting one value
in every one-sided infinite sequences is
\begin{equation}
\left\{
\begin{IEEEeqnarraybox}[][c]{ll}
\IEEEstrut
g_1(S)&=2^{N}\sum_{k=2}^{+\infty}(s^k\cdot 2^{-Nk}),\\
g_2(U)&=2^{N}\sum_{k=2}^{+\infty}(u^k\cdot 2^{-Nk}),\\
&\vdots\\
g_m(V)&=2^{N}\sum_{k=2}^{+\infty}(v^k\cdot 2^{-Nk}).
\IEEEstrut	
\end{IEEEeqnarraybox}
\right.
\label{decimal_fraction2}
\end{equation}

Adding Eq.~(\ref{decimal_integer2}) and (\ref{decimal_fraction2}) together,
one can obtain the general form of the corresponding decimal number,
\begin{equation*}
\left\{
\begin{IEEEeqnarraybox}[\IEEEeqnarraystrutmode\IEEEeqnarraystrutsizeadd{2pt}{2pt}][c]{rCl}
\IEEEeqnarraymulticol{3}{l}{g_1(y_1,y_2,\ldots,y_m)}\\
{\ }& = &(x_1\cdot\overline{s^1})+(F_1(\cdot)\cdot s^1)+2^{N}\sum_{k=2}^{+\infty}(s^k\cdot 2^{-Nk}),\\
\IEEEeqnarraymulticol{3}{l}{g_2(y_1,y_2,\ldots,y_m)}\\
{\ } & = &(x_2\cdot\overline{u^1})+(F_2(\cdot)\cdot u^1)+2^{N}\sum_{k=2}^{+\infty}(u^k\cdot 2^{-Nk}),\\
&\vdots & \\
\IEEEeqnarraymulticol{3}{l}{g_m(y_1,y_2,\ldots,y_m)}\\
{\ } & = & (x_m\cdot\overline{v^1})+(F_m(\cdot)\cdot v^1)+2^{N}\sum_{k=2}^{+\infty}(v^k\cdot 2^{-Nk}),
\end{IEEEeqnarraybox}
\right.
\end{equation*}
after randomly updating multiple random bits and shifting one value in every one-sided infinite sequences.

\subsection{The mathematical expression for $\frac{\partial g_i(y_1,y_2,\ldots,y_m)}{\partial y_j}$}

In the interval $[\frac{n}{2^N}, \frac{n+1}{2^N})$ $(n\in[0,2^{2N}-1])$, the part of decimal integer is not changed that
$\Delta X_1=\Delta X_2= \dots =\Delta X_m=0$. Furthermore,
the first decimals of the $m$ sequences are the same, so
\begin{equation*}
\left\{
\begin{IEEEeqnarraybox}[
\IEEEeqnarraystrutmode
\IEEEeqnarraystrutsizeadd{2pt}
{2pt}
][c]{rCl}
\Delta y_1 & = & \Delta X_1+\Delta S=\Delta S=\sum_{k=2}^{+\infty}(\Delta s^k\cdot 2^{-Nk}),\\
\Delta y_2 & = & \Delta X_2+\Delta U=\Delta U=\sum_{k=2}^{+\infty}(\Delta u^k\cdot 2^{-Nk}),\\
           & \vdots & \\
\Delta y_m & = & \Delta X_m+\Delta U=\Delta U=\sum_{k=2}^{+\infty}(\Delta v^k\cdot 2^{-Nk}).
\end{IEEEeqnarraybox}
\right.
\end{equation*}

From the definition of partial derivative, one has
\begin{equation*}
\begin{IEEEeqnarraybox}[\IEEEeqnarraystrutmode\IEEEeqnarraystrutsizeadd{2pt}{2pt}][c]{rCl}
\IEEEeqnarraymulticol{3}{l}{
\frac{\partial g_1(y_1,y_2,\ldots,y_m)}{\partial y_1} }\\
\quad  & = & \lim_{\Delta y_1\rightarrow 0} \frac{g_1(y_1+\Delta y_1, y_2, \cdots, y_m)-g_1(y_1, y_2, \cdots, y_m)}{\Delta y_1}\\
\quad  & = & \lim_{\Delta S\rightarrow 0}\bigg(\frac{g_1(X)+2^N\sum_{k=2}^{+\infty}((s^k+\Delta s^k)\cdot 2^{-Nk})}{\sum_{k=2}^{+\infty}(\Delta s^k)\cdot 2^{-Nk}} \\
\quad  &   &  \quad -\frac{g_1(X)+2^N\sum_{k=2}^{+\infty}(s^k\cdot 2^{-Nk})}{\sum_{k=2}^{+\infty}(\Delta s^k)\cdot 2^{-Nk}}\bigg)\\
\quad  & = & \lim_{\Delta S\rightarrow 0}\frac{2^N\sum_{k=2}^{+\infty}(\Delta s^k\cdot 2^{-Nk})}{\sum_{k=2}^{+\infty}(\Delta s^k\cdot 2^{-Nk})} \\
\quad  & = & 2^N.
\end{IEEEeqnarraybox}
\end{equation*}
%\enskip, \quad, \qquad leave a horizontal space of respectively half an em, one em and two ems. The "em" is a font depending length, frequently as wide as a %capital M in the current font.

Similarly, one can get
\begin{IEEEeqnarray*}{rCl}
\frac{\partial g_2(y_1, y_2, \cdots, y_m)}{\partial y_2} & = & \frac{\partial g_3(y_1, y_2, \cdots, y_m)}{\partial y_3}\\
                                                         & = & \cdots \\
                                                         & = & \frac{\partial g_m(y_1, y_2, \cdots, y_m)}{\partial y_m}\\
                                                         & = & 2^N.
\end{IEEEeqnarray*}

On the other side, one has
\begin{equation*}
\left\{
\begin{IEEEeqnarraybox}[
\IEEEeqnarraystrutmode
\IEEEeqnarraystrutsizeadd{2pt}
{2pt}
][c]{rCl}
\frac{\partial g_1(y_1, y_2, \cdots, y_m)}{\partial y_2} & = &\frac{\partial g_1(y_1, y_2, \cdots, y_m)}{\partial y_3} = \cdots\\
                                                         & =  & \frac{\partial g_1(y_1, y_2, \cdots, y_m)}{\partial y_m}=0,\\
\frac{\partial g_2(y_1, y_2, \cdots, y_m)}{\partial y_1} & = &\frac{\partial g_2(y_1, y_2, \cdots, y_m)}{\partial y_3}= \cdots  \\
                                                         & =  & \frac{\partial g_2(y_1, y_2, \cdots, y_m)}{\partial y_m}=0,
\\
& \vdots &  \\
\frac{\partial g_m(y_1, y_2, \cdots, y_m)}{\partial y_1} & = &\frac{\partial g_m(y_1, y_2, \cdots, y_m)}{\partial y_2}= \cdots \\
                                                         & = &\frac{\partial g_m(y_1, y_2, \cdots, y_m)}{\partial y_{m-1}}=0.
\end{IEEEeqnarraybox}
\right.
\end{equation*}

So, the corresponding Jacobian matrix, diagonal matrix $\diag(2^N, 2^N, \cdots, 2^N)$,
\iffalse
\begin{equation}
J =
\begin{pmatrix}
2^N & 0 & \ldots & 0 \\
0   & 2^N & \ldots & 0 \\
\vdots   & \vdots & \vdots & \vdots \\
0   & 0 & \ldots & 2^N
\end{pmatrix}
\end{equation}
\fi
is obtained.

\subsection{Estimating the Lyapunov exponents}

Let $\mu_k(\Phi_n^T\cdot \Phi_n)$ denote the $k$-th characteristic value of matrix $(\Phi_n^T\cdot \Phi_n)$, the Lyapunov exponent of the specific
HDDCS can be estimated as
\begin{IEEEeqnarray*}{rCl}
\lambda(y_k) & = & \lim_{n\rightarrow +\infty}\frac{1}{2n}\ln|\mu_k(\Phi_n^T\cdot \Phi_n)| \\
             & = & \lim_{n\rightarrow +\infty}\frac{1}{2n}\ln((2^N)^{2n})\\
            & = & N\ln 2,
\end{IEEEeqnarray*}
where $\Phi_n=J^n$ and $k=1,\ldots,m$ \cite{Li:Lyapunov:C2004}.
	
\section{FPGA-based real-time application of 3D-DCS}
\label{application}

In this section, we first give the design and implementation of an FPGA-based generator for 3D-DCS.
Then, a RGB color image encryption method based on 3D-DCS is presented.
We split the RGB color image into its three R, G, B components, and use the 3D-DCS to scramble the pixel values of the three components
R, G, B. Finally, a systematic methodology for FPGA platform-based implementation of the above method is proposed.

\subsection{Design of 3D-DCS in FPGA}

To utilize the power of FPGA, the computation needs to be divided in several independent blocks of threads that
can be executed simultaneously. The performance on FPGA is directly related to the number of threads and that of logistical elements used during processing, its performances decrease when number of branching instructions (moves like while, if, etc.) increases.
Following these rules, it is possible to build a Verilog-HDL program of the 3D-DCS algorithm \cite{Palnitkar:2003:VHG:1405676}.
	
A concrete example is provided to illustrate the 3D-DCS algorithm. Here, we use 3D-DCS with $N=32$ $(P=32,Q=0)$. For example, a 3-D digital system controlled by random sequences is considered:
\begin{equation}
\left\{
\begin{IEEEeqnarraybox}[][c]{l}
\IEEEstrut
x^n=\overline{x^{n-1}}\oplus(1<\!<(\bmod(z^{n-1},32))),\\
y^n=\overline{y^{n-1}}\oplus(1<\!<(\bmod(x^{n-1},32))),\\
z^n=\overline{z^{n-1}}\oplus(1<\!<(\bmod(y^{n-1},32))).
\IEEEstrut
\end{IEEEeqnarraybox}\right.
\label{3dexp}
\end{equation}
where $\bmod(m, n)=m-n\cdot\lfloor \frac{m}{n} \rfloor$, $\lfloor  \frac{m}{n}\rfloor$ gives the largest integer less than or equal to $ \frac{m}{n}$, and  $<\!<$ represents the left bit shift.
According to Eq.~(\ref{GF}), Eq.~(\ref{BVN3}) and Eq.~(\ref{3dexp}), one can obtain the iterative form of $G_F(E)_{x,y,z}$ as
\begin{equation}
\left\{
\begin{IEEEeqnarraybox}[][c]{l}
\IEEEstrut
x^n=x^{n-1}\!\cdot\!\overline{s^{n}}+((\overline{x^{n-1}}\!\oplus\!(1<\!<(\bmod(z^{n-1},32))))\!\cdot\! s^n),\\
y^n=y^{n-1}\!\cdot\!\overline{u^{n}}+((\overline{y^{n-1}}\!\oplus\!(1<\!<(\bmod(x^{n-1},32))))\!\cdot\! u^n),\\
z^n=z^{n-1}\!\cdot\!\overline{v^{n}}+((\overline{z^{n-1}}\!\oplus\!(1<\!<(\bmod(y^{n-1},32))))\!\cdot\! v^n).
\IEEEstrut	
\end{IEEEeqnarraybox}\right.
\label{gfxyz}
\end{equation}
where $s=s^1s^2s^3\ldots$, $u=u^1u^2u^3\ldots$ and $v=v^1v^2v^3\ldots$ are three random sequences. This 3D-DCS may utilize any reasonable random sequence as $s,u,v$. For demonstration purposes, ISAAC in \cite{robert1996isaac} is adopted here. It was found that the three channels of the outputs of 3D-DCS
all can pass the NIST randomness test suite given in \cite{Rukhin:TestPRNG:NIST10}. Meanwhile, the correlations among the three channels of the outputs
and their auto-correlation strengths are all very low.
		
\begin{figure}[!htb]
\centering
\includegraphics[scale=0.175]{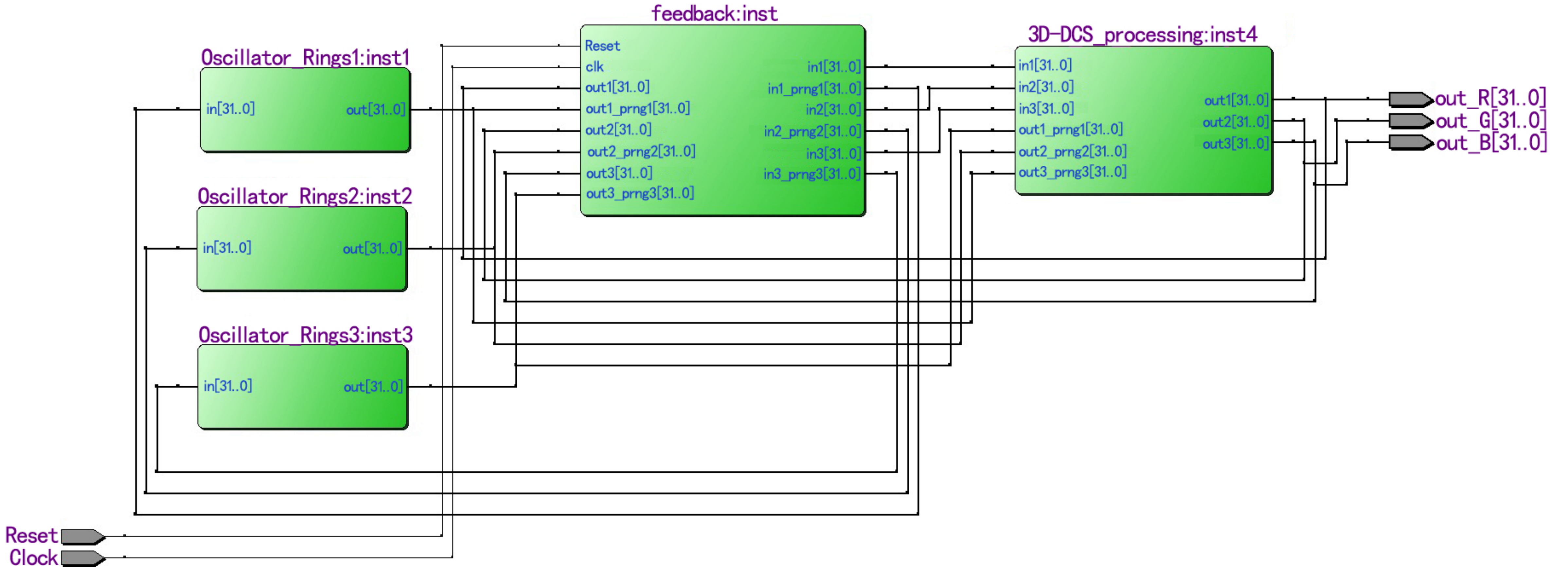}
\caption{Block Diagram of 3D-DCS in FPGA.}
\label{DDCS_3D_Circuit}
\end{figure}

Figure~\ref{DDCS_3D_Circuit} depicts the circuit structure of 3D-DCS. First of all, according to~\cite{Sunar:TRNG:IEEE}, three different types of oscillator rings TRNGs named as \textit{Oscillator\_Rings1:inst1}, \textit{Oscillator\_Rings2:inst2} and \textit{Oscillator\_Rings3:inst3} blocks shown in Fig.~\ref{DDCS_3D_Circuit}, are applied as the external control inputs $s$, $u$, and $v$. Then, the block \textit{3D-DCS\_processing:inst4} is constructed by using Eq.~(\ref{gfxyz}). At last, the states of oscillator rings TRNGs and \textit{3D-DCS\_processing:inst4} are updated in the \textit{feedback} block controlled by clock signal, and it executes in clock positive edge.	
The frequency of input clock decides the speed of the 3D-DCS processing. In our experiments,  the input clock is set as $50$MHz. Then, ModelSim Altera is used to obtain $x^n=out1$, $y^n=out2$ and $z^n=out3$, which are used for RGB image encryption and decryption processing.

\subsection{Design of the FPGA-based hardware system for image encryption and decryption}

\begin{figure}[!htb]
\centering
\includegraphics[scale=0.5]{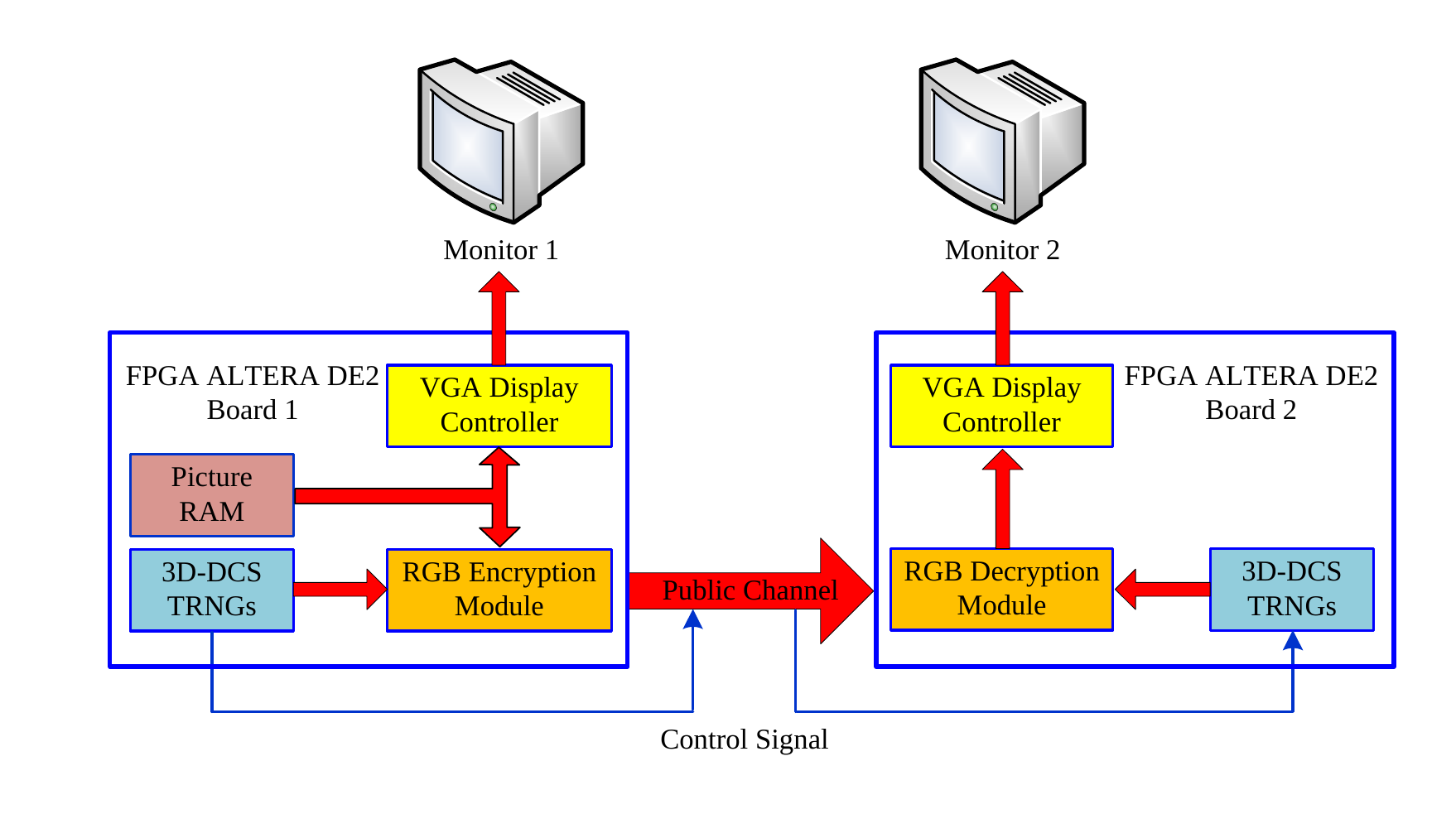}
\caption{Block Diagram of FPGA-Based application on image encryption.}
\label{FPGA Design}
\end{figure}

Figure~\ref{FPGA Design} shows the block diagram for FPGA-based hardware system of image encryption by 3D-DCS,
where the function of the FPGA hardware part is implemented by Verilog-HDL program.

As shown in Fig.~\ref{FPGA Design}, the hardware system at the transmitter side consists of five parts: picture RAM, 3D-DCS, VGA display controller, monitor 1, and RGB encryption module. The picture is encrypted at the transmitter side and
then transmitted through the public channel to the receiver. The receiver, for its part, contains four components: monitor 2, VGA display controller,
RGB decryption module, 3D-DCS.

\begin{figure}[!htb]
\centering
\begin{minipage}{\sfigwidth}
\centering
\includegraphics[width=\sfigwidth]{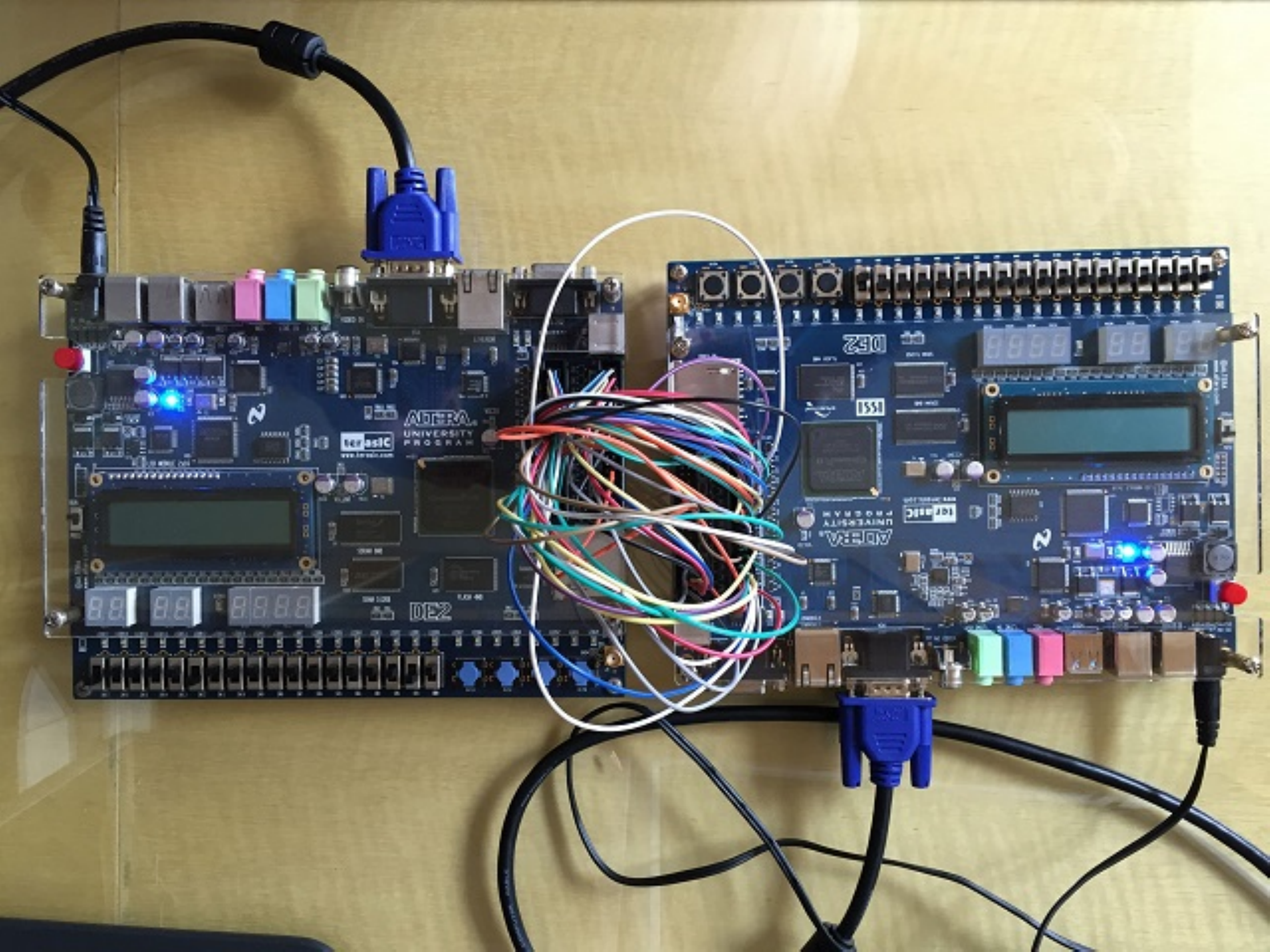}\\
a)
\end{minipage}
\begin{minipage}{\sfigwidth}
\centering
\includegraphics[width=\sfigwidth]{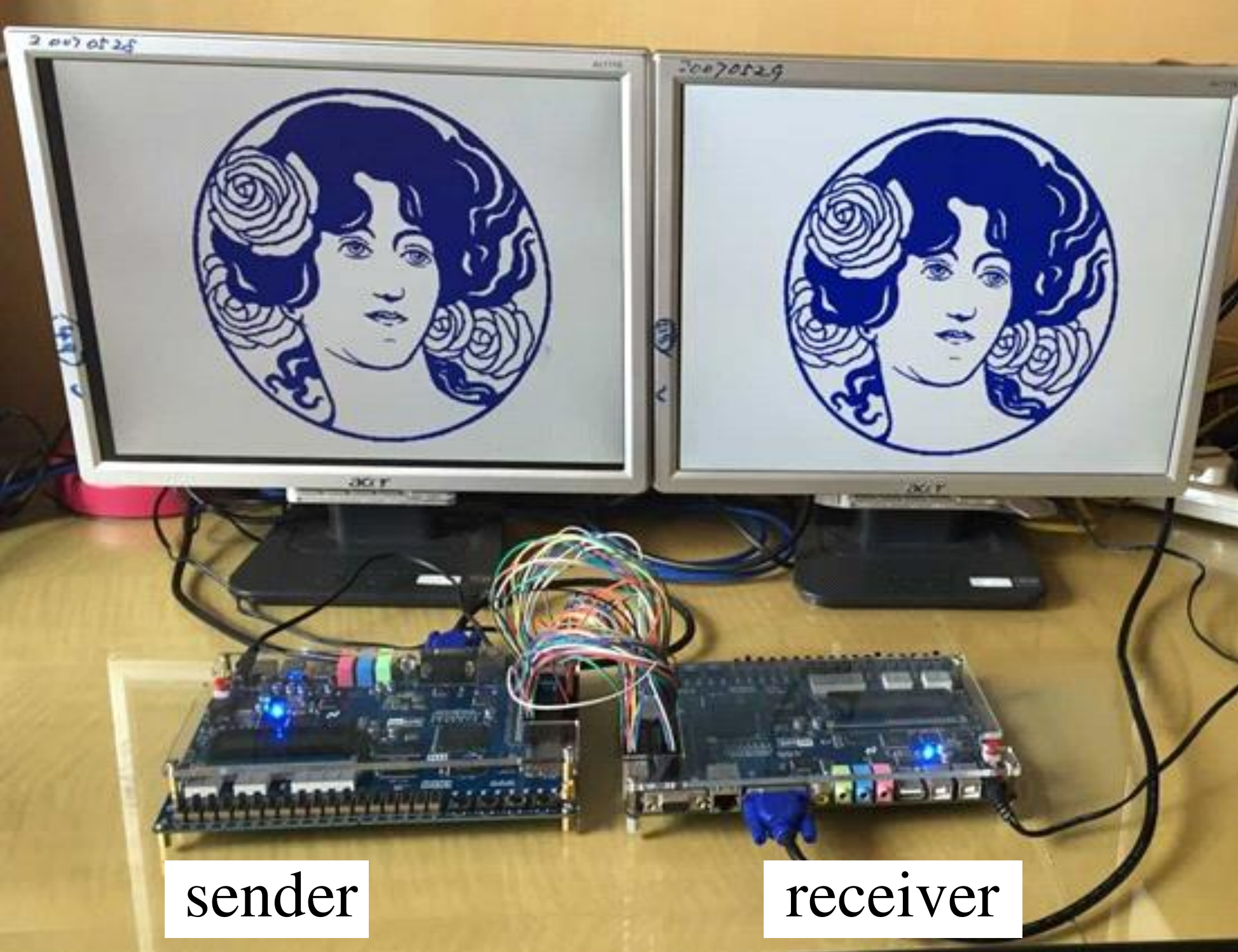}\\
b)
\end{minipage}
\begin{minipage}{\sfigwidth}
\centering
\includegraphics[width=\sfigwidth]{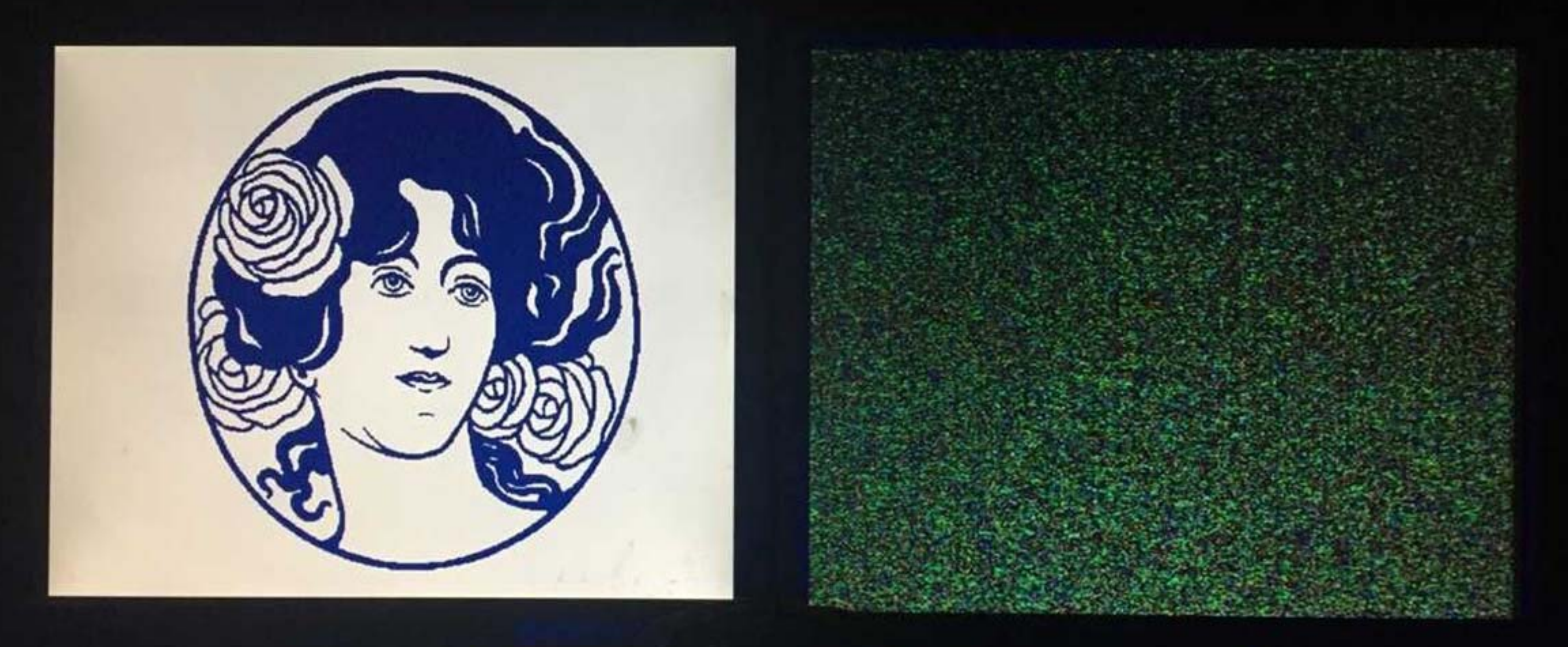}
c)
\end{minipage}
\begin{minipage}{\sfigwidth}
\centering
\includegraphics[width=\sfigwidth]{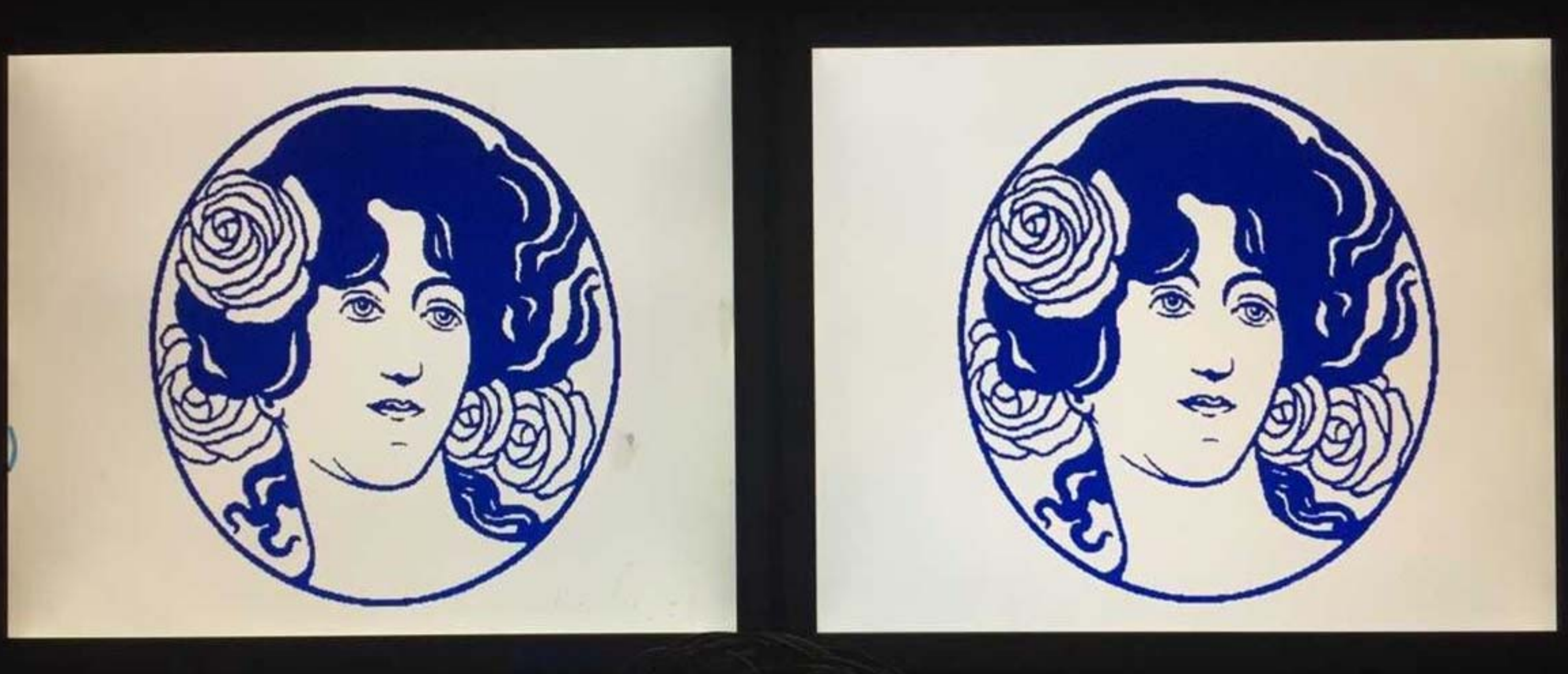}
d)
\end{minipage}
\caption{FPGA-based implementation results for chaos-based secure image communications:
a) Two Altera DE2 FPGA development boards; b) Hardware platform; c) Case of mismatched parameter;
d) Case of matched parameters.}
\label{FPGAimplementation}
\end{figure}

The corresponding hardware implementation platform is shown in Fig.~\ref{FPGAimplementation}, two same models of
Altera DE2 FPGA development board were used. The working principle of the hardware system is expressed as follows.	
In Fig.~\ref{FPGAimplementation}(a), the picture is previously stored in the Altera DE2's RAM (Random Access Memory), namely Picture RAM in Fig.~\ref{FPGA Design}.
When the system is turned on, the picture is sent in two ways: one is delivered to VGA display controller, it collects the RGB (Red, Green, and Blue)
information of every pixel from the picture, then transmits them to monitor for display; the other way is sent to RGB encryption module,
the RGB elements of each pixel are encrypted by operating XOR with the output stream of 3D-DCS, respectively. This needs three different types of oscillator rings as external input sources. In addition, the control signal of states for three oscillator rings at the sender are also delivered through the public channel to synchronize the ones of 3D-DCS at the receiver.
	
In Fig.~\ref{FPGAimplementation}(b), the received encrypted RGB objects are decrypted. Since the states of the 3D-DCS synchronize with the ones
at the transmitter side, the RGB values of the picture could be recovered by executing XOR separately with the states at three
dimensions of 3D-DCS. The RGB pixels are finally sent to VGA controller and displayed on the monitor.

%%%%%%%%%%%%%%%%%%%%%%%%%%%%%%%%%%%%%%%%%%%%%%%%%%%%%%%%%%%%%%%%%%%%%%%%%%%%%%%%%%%%%%%%
\iffalse
\begin{figure*}
\centering%
\subfigure{
\begin{minipage}[b]{0.4\textwidth}
\centering
\includegraphics[width=0.5\linewidth]{twoFPGAdevelopmentboards}

{(a) Two Altera DE2 FPGA development boards}
\end{minipage}}%
\subfigure{
\begin{minipage}[b]{0.4\textwidth}
\centering
\includegraphics[width=0.5\linewidth]{Thehardwareimplementationplatform}

{ (b) Hardware platform}
\end{minipage}}\\ [-5pt]
\subfigure{
\begin{minipage}[b]{0.4\textwidth}
\centering
\includegraphics[width=0.5\linewidth]{Theincorrectlydecryptedimage}

{ (c) Case of mismatched parameter}
\end{minipage}}%
\subfigure{
\begin{minipage}[b]{0.4\textwidth}
\centering
\includegraphics[width=0.5\linewidth]{Thecorrectlydecryptedimage}

{ (d) Case of matched parameters}
\end{minipage}}\\ [-5pt]
\caption{FPGA-based implementation results for chaos-based secure image communications.}
\label{fig7}%original fig4
\end{figure*}
\fi

\subsection{FPGA-based implementation result for image encryption and decryption}

In the experiment, an image of $640 \times 480$ resolution is previously stored in Altera DE2 board's RAM.
The image encryption system based on 3D-DCS is implemented with this image.
As shown in Fig.~\ref{FPGA Design}, only when the states of 3D-DCS between the transmitter and receiver are exactly matched,
then the original image can be correctly decrypted, otherwise it cannot be recovered. The experimental FPGA-based implementation results are
shown in Figs.~\ref{FPGAimplementation}(c) and (d).
	
\section{Conclusion}

To solve the degradation of chaotic dynamical properties caused by limitation of finite-precision presentation and quantization,
this paper developed a novel higher dimensional digital chaotic system (HDDCS), utilizing the chaos generation strategy
controlled by random sequences. It is proved that HDDCS satisfies the Devaney¡¯s definition of chaos. Finally, FPGA-based
implementation of image cryptography algorithm with 3D-DCS is detailed, confirming the feasibility and applicability.
This work provides perfect solution to dynamical degradation of digital chaos and may further promote application of chaos
in digital world.

%\bibliographystyle{IEEEtran}
%\bibliography{HDDCS}

\begin{thebibliography}{99}
\bibitem{MAY:Logistic:Nature1976}
R.~M. May, ``Simple mathematical models with very complicated dynamics,''
  \emph{Nature}, vol. 261, no. 5560, pp. 459--467, 1976.

\bibitem{YaobinMao:CSF2004}
G.~Chen, Y.~Mao, and C.~K. Chui, ``A symmetric image encryption scheme based on
  {3D} chaotic cat maps,'' \emph{Chaos, Solitons \& Fractals}, vol.~21, no.~3,
  pp. 749--761, 2004.

\bibitem{LiShujun:Rules:IJBC2006}
G.~Alvarez and S.~Li, ``Some basic cryptographic requirements for chaos-based
  cryptosystems,'' \emph{International Journal of Bifurcation and Chaos},
  vol.~16, no.~8, pp. 2129--2151, 2006.

\bibitem{kanso2009logistic}
A.~Kanso and N.~Smaoui, ``Logistic chaotic maps for binary numbers
  generations,'' \emph{Chaos, Solitons {\&} Fractals}, vol.~40, no.~5, pp.
  2557--2568, 2009.

\bibitem{Kohda:Statistics:TIT1997}
T.~Kohda and A.~Tsuneda, ``Statistics of chaotic binary sequences,'' \emph{IEEE
  Transactions on Information Theory}, vol.~43, no.~1, pp. 104--112, Jan 1997.

\bibitem{Gregory:Commun:Science98}
G.~D. VanWiggeren and R.~Roy, ``Communication with chaotic lasers,''
  \emph{Science}, vol. 279, no. 5354, pp. 1198--1200, 1998.

\bibitem{Lee:lyapunov:CPC2004}
P.-H. Lee, Y.~Chen, S.-C. Pei, and Y.-Y. Chen, ``Evidence of the correlation
  between positive lyapunov exponents and good chaotic random number
  sequences,'' \emph{Computer Physics Communications}, vol. 160, no.~3, pp.
  187--203, 2004.

\bibitem{Kwok:finiteprecision:2007}
H.~S. Kwok and W.~K.~S. Tang, ``A fast image encryption system based on chaotic
  maps with finite precision representation,'' \emph{Chaos Solitons {\&}
  Fractals}, vol.~32, no.~4, pp. 1518--1529, 2007.

\bibitem{Cho:QKD:TCASI15}
K.~Cho and T.~Miyano, ``Chaotic cryptography using augmented lorenz equations
  aided by quantum key distribution,'' \emph{IEEE Transactions on Circuits and
  Systems I: Regular Papers}, vol.~62, no.~2, pp. 478--487, Feb 2015.

\bibitem{beck1987effects}
C.~Beck and G.~Roepstorff, ``Effects of phase space discretization on the
  long-time behavior of dynamical systems,'' \emph{Physica D}, vol.~25, no.~1,
  pp. 173--180, 1987.

\bibitem{blank1994pathologies}
M.~Blank, ``Pathologies generated by round-off in dynamical systems,''
  \emph{Physica D}, vol.~78, no.~1, pp. 93--114, 1994.

\bibitem{Binder:Logistic:PRA86}
P.~M. Binder and R.~V. Jensen, ``Simulating chaotic behavior with finite-state
  machines,'' \emph{Physical Review A}, vol.~34, no.~5, pp. 4460--4463, 1986.

\bibitem{Lisj:precision:CPC2003}
S.~Li, X.~Mou, Y.~Cai, Z.~Ji, and J.~Zhang, ``On the security of a chaotic
  encryption scheme: Problems with computerized chaos in finite computing
  precision,'' \emph{Computer Physics Communications}, vol. 153, no.~1, pp.
  52--58, 2003.

\bibitem{HPhu:DCS:SMCS2015}
Y.~Deng, H.~Hu, W.~Xiong, N.~N. Xiong, and L.~Liu, ``Analysis and design of
  digital chaotic systems with desirable performance via feedback control,''
  \emph{IEEE Transactions on Systems, Man, and Cybernetics: Systems}, vol.~45,
  no.~8, pp. 1187--1200, Aug 2015.

\bibitem{Galias:Round:CASM13}
Z.~Galias, ``The dangers of rounding errors for simulations and analysis of
  nonlinear circuits and systems--and how to avoid them,'' \emph{IEEE Circuits
  and Systems Magazine}, vol.~13, no.~3, pp. 35--52, Aug 2013.

\bibitem{Deng:hybrid:IS2015}
Y.~Deng, H.~Hu, N.~Xiong, W.~Xiong, and L.~Liu, ``A general hybrid model for
  chaos robust synchronization and degradation reduction,'' \emph{Information
  Sciences}, vol. 305, pp. 146--164, 2015.

\bibitem{hong1997realizing}
H.~Zhou and X.~Ling, ``Realizing finite precision chaotic systems via
  perturbation of $m$-sequences,'' \emph{Acta Electronica Sinica}, vol.~7, p.
  020, 1997.

\bibitem{tao1998perturbance}
T.~Sang, R.~Wang, and Y.~Yan, ``Perturbance-based algorithm to expand cycle
  length of chaotic key stream,'' \emph{Electronics Letters}, vol.~34, no.~9,
  pp. 873--874, 1998.

\bibitem{Cernak:DCS:PLA1996}
J.~Cernak, ``Digital generators of chaos,'' \emph{Physics Letters A}, vol. 214,
  no. 3-4, pp. 151--160, 1996.

\bibitem{Sang:Perturbance:EL1998}
T.~Sang, R.~Wang, and Y.~Yan, ``Perturbance-based algorithm to expand cycle
  length of chaotic key stream,'' \emph{Electronics Letters}, vol.~34, no.~9,
  pp. 873--874, 1998.

\bibitem{LiCY:PRNS:VLSI2012}
C.-Y. Li, Y.-H. Chen, T.-Y. Chang, L.-Y. Deng, and K.~To, ``Period extension
  and randomness enhancement using high-throughput reseeding-mixing prng,''
  \emph{IEEE Transactions on Very Large Scale Integration ({VLSI}) Systems},
  vol.~20, no.~2, pp. 385--389, Feb 2012.

\bibitem{wolff1986transients}
W.~Wolff and B.~Huberman, ``Transients and asymptotics in granular phase
  space,'' \emph{Zeitschrift f{\"u}r Physik B Condensed Matter}, vol.~63,
  no.~3, pp. 397--405, 1986.

\bibitem{Nagaraj:EPJT:2008}
N.~Nagaraj, M.~C. Shastry, and P.~G. Vaidya, ``Increasing average period
  lengths by switching of robust chaos maps in infinite precision,''
  \emph{European Physical Journal-Special Topics}, vol. 165, no.~1, pp. 73--83,
  2008.

\bibitem{YCZhou:Switching:TCASI2014}
Y.~Wu, Y.~Zhou, and L.~Bao, ``Discrete wheel-switching chaotic system and
  applications,'' \emph{IEEE Transactions on Circuits and Systems I-Regular
  Papers}, vol.~61, no.~12, pp. 3469--3477, Dec 2014.

\bibitem{Li:logistic:ND2014}
C.~Li, T.~Xie, Q.~Liu, and G.~Cheng, ``Cryptanalyzing image encryption using
  chaotic logistic map,'' \emph{Nonlinear Dynamics}, vol.~78, no.~2, pp.
  1545--1551, 2014.

\bibitem{Li:DPWLCM:IJBC2005}
S.~Li, G.~Chen, and X.~Mou, ``On the dynamical degradation of digital piecewise
  linear chaotic maps,'' \emph{International Journal of Bifurcation and Chaos},
  vol.~15, no.~10, pp. 3119--3151, 2005.

\bibitem{Guyeux:Hash:JACT10}
C.~Guyeux and J.~Bahi, ``Hash functions using chaotic iterations,''
  \emph{Journal of Algorithms \& Computational Technology}, vol.~4, no.~2, pp.
  167--182, 2010.

\bibitem{Devaney:Chaos:2003}
R.~L. Devaney, \emph{An introduction to chaotic dynamical systems}.\hskip 1em
  plus 0.5em minus 0.4em\relax Boulder, Colorado, USA: Westview Press, 2003.

\bibitem{Bahi:XORshift:JNCA04}
J.~M. Bahi, X.~Fang, C.~Guyeux, and Q.~Wang, ``Suitability of chaotic
  iterations schemes using xorshift for security applications,'' \emph{Journal
  of Network and Computer Applications}, vol.~37, pp. 282--292, 2014.

\bibitem{Bahi:PRNS:IJAS11}
------, ``Evaluating quality of chaotic pseudo-random generators. application
  to information hiding,'' \emph{International Journal on Advances in
  Security}, vol.~4, no. 1-2, pp. 118--130, 2011.

\bibitem{SMYu:Chaotifying:IJBC2012}
S.~Yu and G.~Chen, ``Chaotifying continuous-time nonlinear autonomous
  systems,'' \emph{International Journal of Bifurcation and Chaos}, vol.~22,
  no.~9, p. art. no. 1250232, 2012.

\bibitem{SMYu:integer:IJBC2014}
Q.~Wang, S.~Yu, C.~Guyeux, J.~Bahi, and X.~Fang, ``Theoretical design and
  circuit implementation of integer domain chaotic systems,''
  \emph{International Journal of Bifurcation and Chaos}, vol.~24, no.~10, p.
  Art. no. 1450128, 2014.

\bibitem{SYu:ARM:CASVT15}
Z.~Lin, S.~Yu, J.~Lu, S.~Cai, and G.~Chen, ``Design and {ARM}-embedded
  implementation of a chaotic map-based real-time secure video communication
  system,'' \emph{IEEE Transactions on Circuits and Systems for Video
  Technology}, vol.~25, no.~7, pp. 1203--1216, Jul 2015.

\bibitem{Wang:bitwise:CPB15}
Q.~Wang, S.~Yu, C.~Guyeux, J.~Bahi, and X.~Fang, ``Study on a new chaotic
  bitwise dynamical system and its {FPGA} implementation,'' \emph{Chinese
  Physics B}, vol.~24, no.~6, p. 60503, 2015.

\bibitem{hirano2010fast}
K.~Hirano and et~al., ``Fast random bit generation with bandwidth-enhanced
  chaos in semiconductor lasers,'' \emph{Optics express}, vol.~18, no.~6, pp.
  5512--5524, 2010.

\bibitem{Kocarev:TrueRandom:TCASI11}
T.~Addabbo, A.~Fort, L.~Kocarev, S.~Rocchi, and V.~Vignoli, ``Pseudo-chaotic
  lossy compressors for true random number generation,'' \emph{IEEE
  Transactions on Circuits and Systems I: Regular Papers}, vol.~58, no.~8, pp.
  1897--1909, Aug 2011.

\bibitem{Michael:TRNG:Lecture}
M.~Epstein, L.~Hars, R.~Krasinski, M.~Rosner, and H.~Zheng, ``Design and
  implementation of a true random number generator based on digital circuit
  artifacts,'' \emph{Lecture Notes in Computer Science}, vol. 2779, pp.
  152--165, 2003.

\bibitem{Sunar:TRNG:IEEE}
B.~Sunar, W.~Martin, and D.~Stinson, ``A provably secure true random number
  generator with built-in tolerance to active attacks,'' \emph{IEEE
  Transactions on Computers}, vol.~56, no.~1, pp. 109--119, Jan 2007.

\bibitem{Hsu:CellMap:IJBC92}
C.-S. Hsu, ``Global analysis by cell mapping,'' \emph{International Journal of
  Bifurcation and Chaos}, vol.~2, no.~4, pp. 727--771, 1992.

\bibitem{Shreim:NetworkCA:2007}
A.~Shreim, P.~Grassberger, W.~Nadler, B.~Samuelsson, J.~E.~S. Socolar, and
  M.~Paczuski, ``Network analysis of the state space of discrete dynamical
  systems,'' \emph{Physical Review Letters}, vol.~98, no.~19, p. 198701, 2007.

\bibitem{Banks:definition:AMM1992}
J.~Banks, J.~Brooks, G.~Cairns, G.~Davis, and P.~Stacey, ``On devaney's
  definition of chaos,'' \emph{American Mathematical Monthly}, vol.~99, no.~4,
  pp. 332--334, 1992.

\bibitem{Li:Lyapunov:C2004}
C.~Lin and G.~Chen, ``Estimating the lyapunov exponents of discrete systems,''
  \emph{Chaos}, vol.~14, pp. 343--346, Jun 2004.

\bibitem{Palnitkar:2003:VHG:1405676}
S.~Palnitkar, \emph{Verilog HDL: a guide to digital design and synthesis},
  2nd~ed.\hskip 1em plus 0.5em minus 0.4em\relax California, USA: Prentice Hall
  Professional, 2003.

\bibitem{robert1996isaac}
R.~J.~J. Jenkins, ``Isaac,'' \emph{Fast Software Encryption}, vol. 1039, pp.
  41--49, 1996.

\bibitem{Rukhin:TestPRNG:NIST10}
A.~Rukhin and et~al., ``A statistical test suite for random and pseudorandom
  number generators for cryptographic applications,'' NIST Special Publication
  800-22rev1a, 2010, available online at
  \url{http://csrc.nist.gov/groups/ST/toolkit/rng/documentation_software.html}.
\end{thebibliography}

%\iffalse
%%%%%%%%%%%%%%%%%%%%%%%%%%%%%%%%%%%%%%%%%%%%%%%%%%%%%%%%%%%%%%%%%%%%%%%%%%%%%%%%%%%%%%%%%%%%%%%%%%%%%%%%%%%%%%%%
\graphicspath{{author-figures-pdf/}}

%\vspace{-10mm}

\begin{IEEEbiography}[{\includegraphics[width=1in,height=1.25in,clip,keepaspectratio]{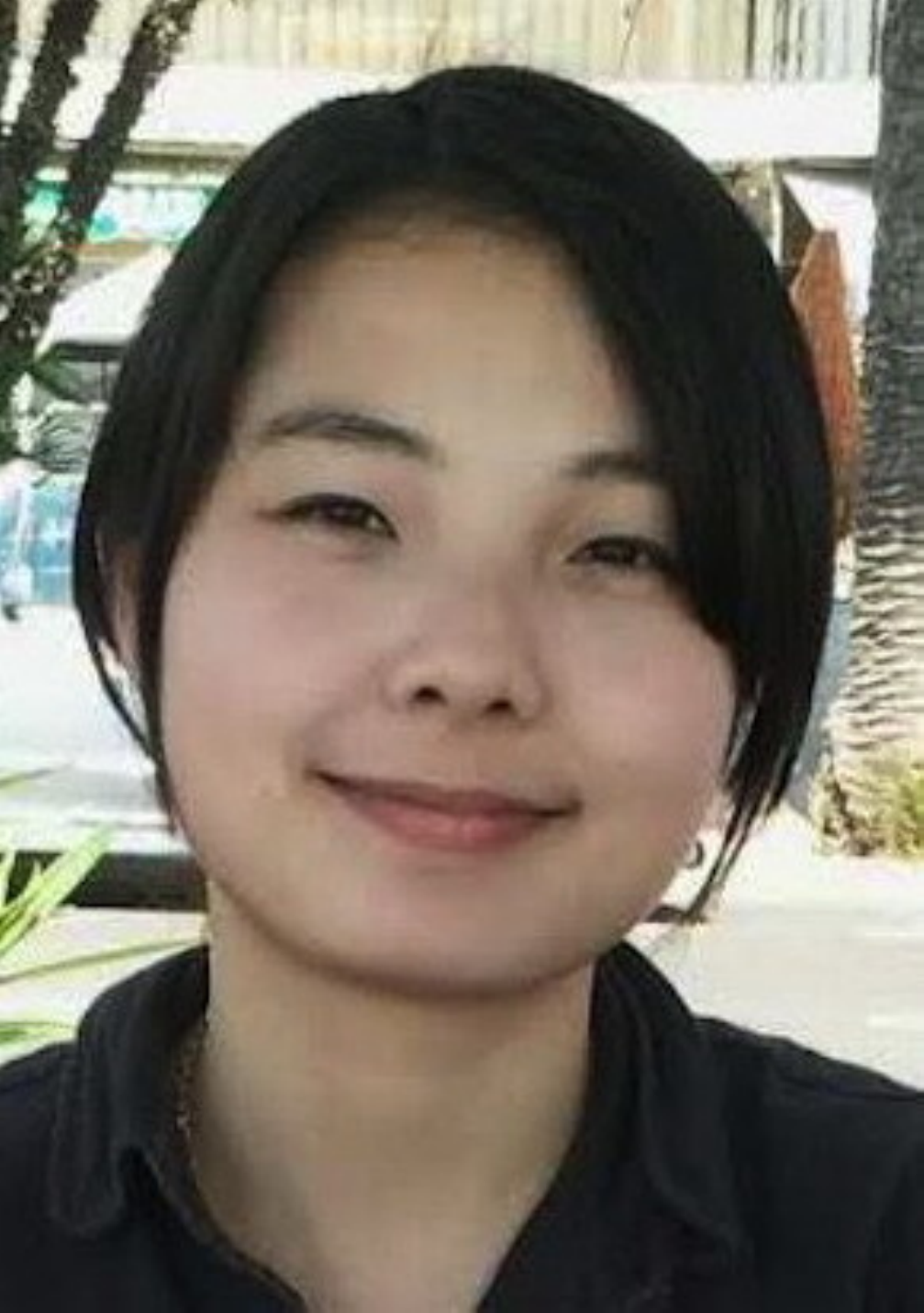}}]{Qianxue Wang}
received the Ph.D. degree in computer science and engineering, from the Department of Complex System (DISC), FEMTO-ST Institute, University of Franche-Comt¨¦, supervised by Prof. Jacques Bahi. In 2013, she started a postdoctoral position in Guangdong University of Technology where she is currently working on chaotic systems with Prof. Simin Yu. Since 2008, she has published 5 articles in international journals, and 11 articles in peer reviewed international conferences dealing with chaos.
\end{IEEEbiography}

\vspace{-15mm}

\begin{IEEEbiography}[{\includegraphics[width=1in,height=1.25in,clip,keepaspectratio]{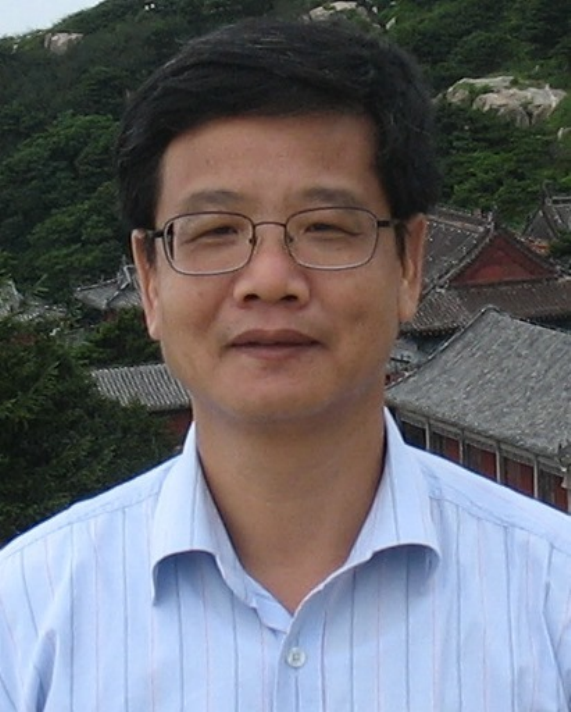}}]{Simin Yu}
received the B.Sc. degree in electronics from Yunnan University, Kunming, China in 1983,
the M.E. degree in radio communication engineering, and the Ph.D. degree in nonlinear circuits and
systems, both from the South China University of Technology, Guangzhou, China in 1996 and 2001,
respectively. Currently, he is a Professor in the College of Automation, Guangdong University of Technology,
China. His research interests include design and analysis of nonlinear circuits, and wireless communications. He has published more than 60
SCI journal papers, 3 academic books and 6 granted patents in the fields
of chaos in electronic circuits and chaotic communications.
\end{IEEEbiography}

\vspace{-10mm}

\begin{IEEEbiography}[{\includegraphics[width=1in,height=1.25in,clip,keepaspectratio]{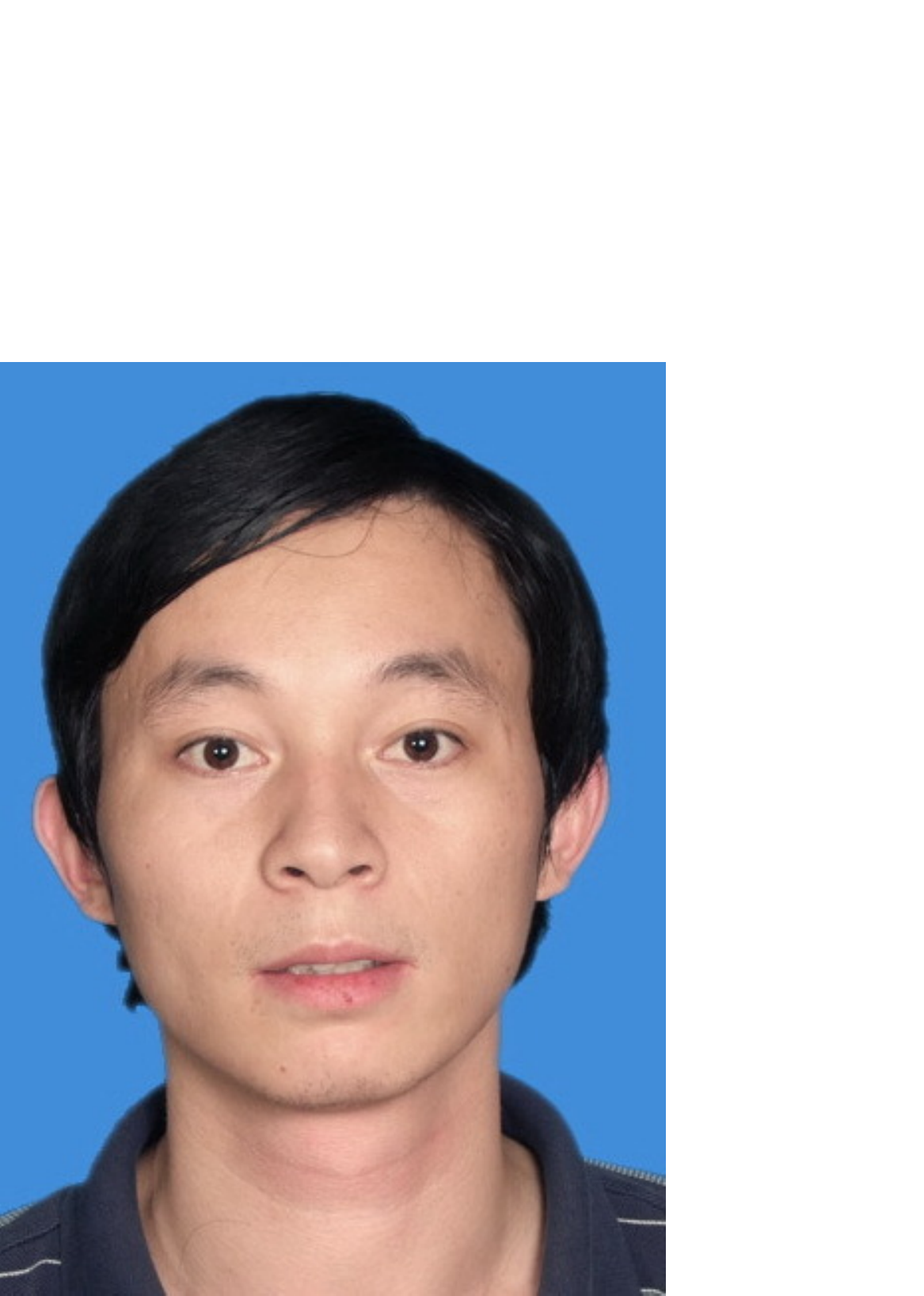}}]{Chengqing Li}(M'07-SM'13)
received his M.Sc. degree in Applied Mathematics from Zhejiang University, China in 2005 and his Ph.D. degree in Electronic Engineering from City University of Hong Kong in 2008. Thereafter, he had been working as a Postdoctoral Fellow at The Hong Kong Polytechnic University till September 2010. Then, he joined the College of Information Engineering, Xiangtan University, China as an Associate Professor, where he received his Bachelor degree in Mathematics before. From April 2013 to July 2014, he worked at University of Konstanz, Germany, under the support of the Alexander von Humboldt Foundation.
He is serving as an associate editor for International Journal of Bifurcation and Chaos. Dr. Li focuses on security analysis of image and chaos-based encryption schemes.
\end{IEEEbiography}

\vspace{-10mm}

\begin{IEEEbiography}[{\includegraphics[width=1in,height=1.25in,clip,keepaspectratio]{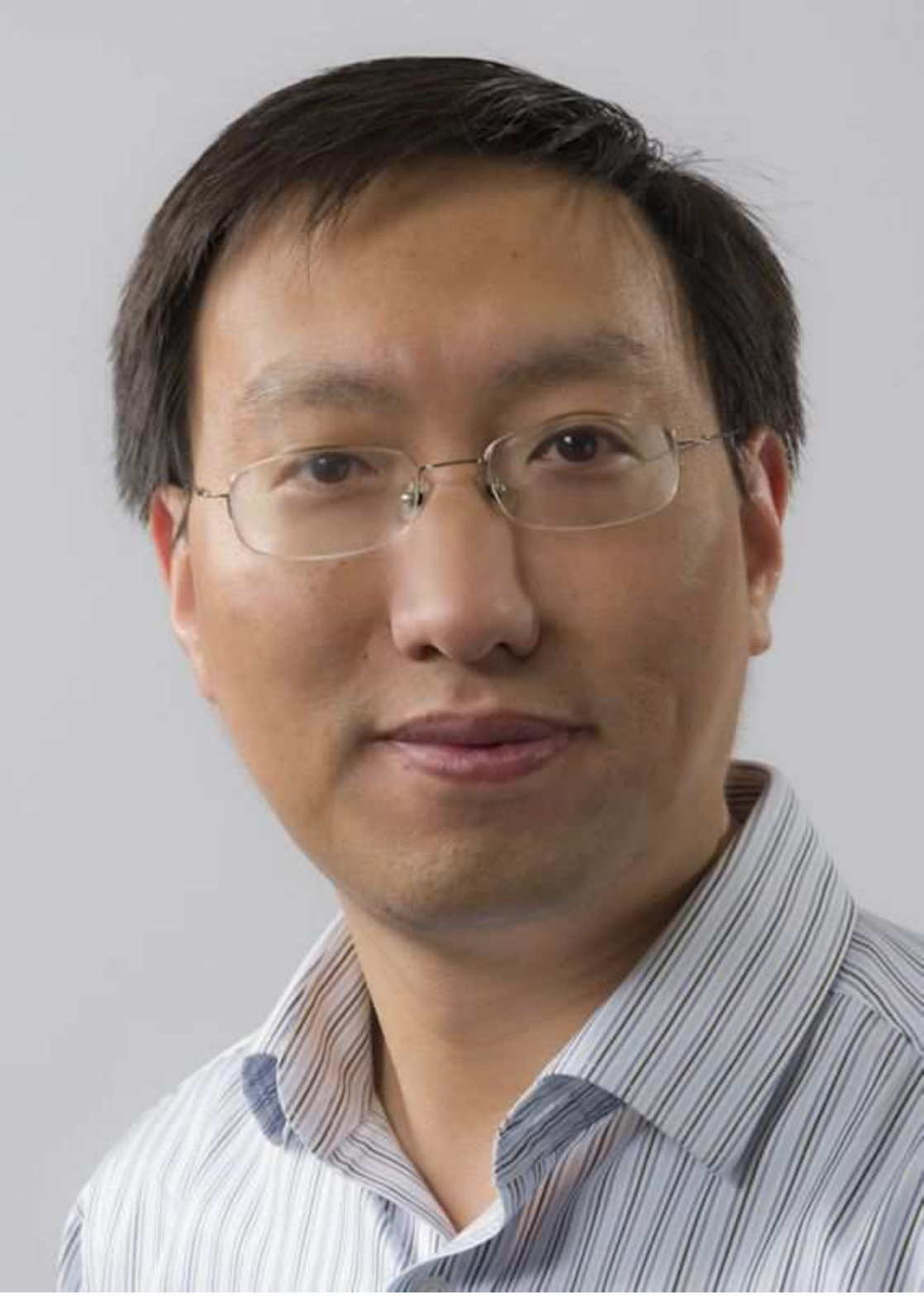}}]{Jinhu L\"u}(M'03-SM'06-F'13)
received the Ph.D degree in applied mathematics from the Academy of Mathematics and Systems Science, Chinese Academy
of Sciences, Beijing, China in 2002. Currently, he is a Professor in the Academy of Mathematics and
Systems Science, Chinese Academy of Sciences. He was a Professor in the RMIT University, Australia
and a Visiting Fellow in the Princeton University, USA. He is the author of three research monographs
and more than 110 SCI journal papers published in the fields of complex networks and complex systems,
nonlinear circuits and systems, receiving more than 8000 SCI citations with h-index 42. He is an ISI Highly Cited Researcher in Engineering.
\end{IEEEbiography}

\vspace{-10mm}

\begin{IEEEbiography}[{\includegraphics[width=1in,height=1.25in,clip,keepaspectratio]{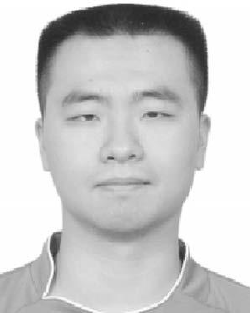}}]{Xiaole Fang}
received the Ph.D. degree in April 2013 in computer science and engineering, from the Department of Complex System (DISC), FEMTO-ST Institute, University of Franche-Comt¨¦, supervised by Prof. Jacques Bahi and Prof. Laurent Larger. Recently, he is working as a GIS (Geographic Information System) engineer in Land and Resources Technology Center, Guangdong Province, China. Since 2010, he has published seven articles in international journal, and seven articles in peer reviewed international conferences.
\end{IEEEbiography}

\vspace{-10mm}

\begin{IEEEbiography}[{\includegraphics[width=1in,height=1.25in,clip,keepaspectratio]{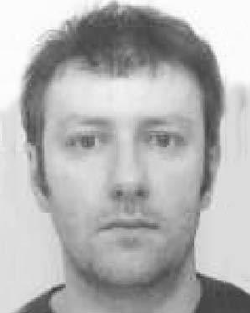}}]{Christophe Guyeux}
has taught mathematics and computer science in the Belfort-Montb¨¦liard University Institute of Technologies (IUT-BM) this last decade. He has defended a computer science thesis dealing with security, chaos, and dynamical systems in 2010 under Jacques Bahi¡¯s leadership, and is now an associated professor in the computer science department of complex system (DISC), FEMTO-ST Institute, University of Franche-Comt\'{e}. Since 2010, he has published 2 books, 9 articles in international journals, and 25 articles in peer reviewed international conferences dealing with security or chaos.
\end{IEEEbiography}

\vspace{-10mm}

\begin{IEEEbiography}[{\includegraphics[width=1in,height=1.25in,clip,keepaspectratio]{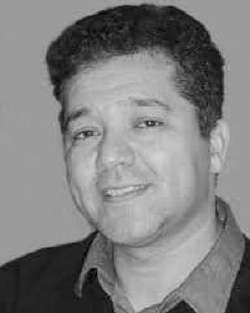}}]{Jacques M. Bahi}
received Ph.D. degrees in applied mathematics from the University of Franche-Comte, France, in 1991. From 1992 to 1999, he was an Associate Professor of applied mathematics at the Mathematical Laboratory of Besancon. His research interests were focused on parallel synchronous and asynchronous algorithms for differential-algebraic equations and singular systems. Since September 1999, he has been a Full Professor of computer science at the University of Franche-Comte. He published about 150 articles in peer reviewed journal and international conferences and 2 scientific books. He is a member of the editorial board of 2 international journals and is regularly a member of the scientific commitees of many international conferences. Currently, he is interested in 1) high performance computing, 2) distributed numerical algorithms for ad-hoc and sensor networks and 3) dynamical systems with application to data hiding and privacy.
\end{IEEEbiography}
%\fi
\end{document}